\documentclass[%
 reprint,
 superscriptaddress,
 amsmath,amssymb,
 aps,pra]{revtex4-2}

\usepackage{graphicx}
\usepackage{dcolumn}
\usepackage{bm}
\usepackage[version=4]{mhchem}
\usepackage{physics}
\usepackage{multirow}
\usepackage{siunitx}
\usepackage[dvipsnames]{xcolor}
\usepackage{hyperref}

\newcommand{\Modefunc}{\boldsymbol{\mathsf{M}}}

\begin{document}
\title{On intermolecular interactions in the Hamiltonian used in polaritonic chemistry}

\author{{Marit R. Fiechter}}
\email{marit.fiechter@phys.chem.ethz.ch}
\affiliation{Department of Chemistry and Applied Biosciences, ETH Z\"urich, 8093 Z\"urich, Switzerland}

\author{Mark Kamper Svendsen}
\affiliation{NNF Quantum Computing Programme, Niels Bohr Institute, University of Copenhagen, Universitetsparken 5, 2100 Copenhagen, Denmark}

\date{\today}
\begin{abstract} 
Experiments have shown that strong coupling between molecular excitations and a mode of a Fabry--P\'erot cavity can significantly alter molecular properties, such as reaction rates and equilibrium constants. However, in spite of the large body of theoretical work, the mechanism behind this change is still not well understood. 
In order to make progress, we first take a step back and investigate the appropriateness of the Hamiltonian that most recent studies are based on. In particular, we investigate the dipole self-energy, which can be divided into in self terms and cross terms. While the self terms are an indispensable part of the Hamiltonian, the cross terms -- which have received attention as they seem to mediate distance-independent interactions between all molecules in the cavity --  
are known to, under certain conditions, cancel exactly with the usually neglected intermolecular Coulombic interactions. 
In this work, we revisit how this cancellation comes about in free space and in a perfect cavity, clarifying that it can only be found when looking beyond the single-mode approximation and taking the full continuum of light modes into account. 
We also provide numerical evidence suggesting that this cancellation may extend to the case of an imperfect cavity, and show how the situation changes for a more realistic cavity in the framework of macroscopic QED. Finally, we discuss the implications of this cancellation for the single-mode Hamiltonian.   
\end{abstract} 
\maketitle 

\section{Introduction} 
\begin{figure*}
    \centering
    \includegraphics[width=0.9\linewidth]{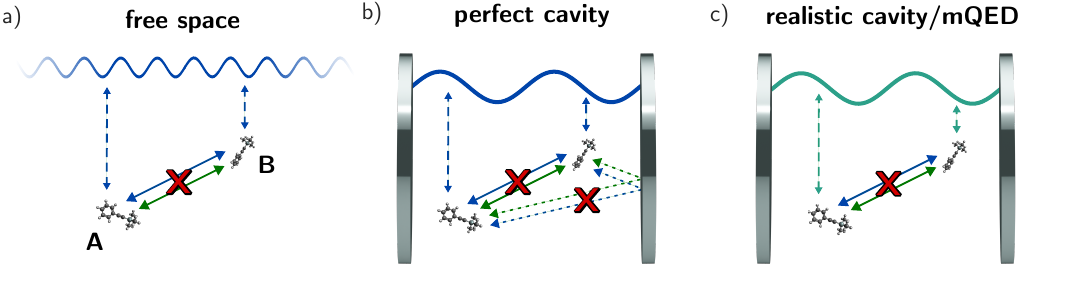}
    \caption{Schematic summary of the three cases considered in this work. a) Free space: the direct interactions (Coulombic dipole--dipole interaction in solid green, and DSE cross term in solid blue) exactly cancel {under certain conditions}, and the molecules only interact with each other via the displacement field. b) Perfect cavity: just like in free space, the direct interactions between the molecules cancel; additional cavity-mediated Coulombic (dotted green) and DSE (dotted blue) interactions arise in this case (interactions with the second mirror not shown), and exactly cancel each other out too. Again, all interactions are mediated by the displacement field, but the mode structure of this field is now modified by the cavity. c) Imperfect cavity in a macroscopic QED framework: as fluctuations in the mirror material can affect the field in the cavity, the field is now most conveniently described in terms of mixed light--matter operators. As before, the free-space-like direct interactions cancel; the additional cavity-mediated Coulombic and DSE interactions are also affected by the fluctuations in the mirrors, and are therefore naturally absorbed in the matter-field interaction term (dashed teal arrows).  }
    \label{fig:summary}
\end{figure*}

Recent experiments have demonstrated that one can modify chemistry by simply placing a pair of mirrors around a reaction mixture, thereby forming a Fabry--P\'erot microcavity. For example, strongly coupling a molecular vibration to a cavity mode \cite{nagarajan2021chemistry,dunkelberger2022vibration} can lead to a change in chemical properties such as thermal reaction rates \cite{thomas2016ground,thomas2019tilting,lather2019cavity,vergauwe2019modification,hirai2020modulation,sau2021modifying,ahn2023modification}, equilibrium constants \cite{pang2020role, patrahau2024direct} and self-assembly products \cite{sandeep2022manipulating,imai2025accessing}. This has opened up the field of polaritonic chemistry, which brings with it a realm of new possibilities: for example, one could envision harnessing this effect and using a cavity to selectively promote a desired reaction while suppressing side reactions, so that the cavity effectively acts as a non-invasive catalyst \cite{thomas2019tilting, george2023polaritonic}. 

Because of this exciting prospect, a lot of theoretical work has gone into understanding the mechanism behind the observed change in rate \cite{campos2023swinging,mandal2023theoretical,ruggenthaler2023understanding,galego2019cavity,li2020origin,li2021cavity,yang2021quantum,lindoy2022resonant,wang2022cavity,wang2022chemical,philbin2022chemical,sun2022suppression,schafer2022shining,du2023vibropolaritonic,lindoy2023quantum,fiechter2023RPMD,sun2023modification,anderson2023mechanism, sidler2024unraveling,ying2023resonance, lindoy2024investigating,ke2024insights,sidler2024connection,ying2024resonance,ke2025stochastic,horak2025analytic,vega2025theoretical}. However, so far no conclusive explanation has been found. The most promising hypothesis has been that of rate enhancement by modification of vibrational excitation and de-excitation rates \cite{lindoy2023quantum, lindoy2024investigating,fiechter2023RPMD,ke2024insights, ke2025stochastic, ying2023resonance, ying2024resonance}, as first proposed by Lindoy \emph{et al.} based on full quantum-dynamical calculations for a model system \cite{lindoy2023quantum}. 
As is currently standard in the field, the light--matter coupling in this study is described by
\begin{equation}\label{eq:singlemode}
    \hat{V}_\mathrm{cav} = \tfrac{1}{2}\omega_\mathrm{c}^2\bigg(\hat{q}_\mathrm{c} +\sqrt{\frac{2}{\omega_\mathrm{c}}}\eta \hat{\mu}_g \bigg)^2,
\end{equation}
where the molecule is coupled to a single cavity mode $\hat{q}_\mathrm{c}$ of frequency $\omega_\mathrm{c}$, with coupling strength $\eta$, via its dipole moment {in the uncoupled electronic ground state} $\hat{\mu}_g$.
Currently, this mechanism is the only one proposed that can reproduce the sharp resonance feature in rate modification at cavity frequencies that match a vibrational frequency, as is observed in experiment. 
However, 
when increasing the number of molecules and decreasing the coupling strength to realistic values, the effect washes out \cite{lindoy2024investigating,ke2025stochastic} -- meaning that this approach predicts no observable changes in the reaction rate under collective strong coupling, which is contrary to what is seen in experiment. 
With this finding, it now seems clear that the widely-used ground-state single-mode Hamiltonian in Eq.~\eqref{eq:singlemode} 
 is insufficient to explain the experiments. 
This has prompted us to re-investigate the form of the Hamiltonian used in rate studies.

In recent years, the presence of the dipole self-energy (DSE) term in the Hamiltonian in the electric-dipole gauge (\emph{i.e.} the $\hat{\mu}^2$--term in Eq.~\eqref{eq:singlemode}) has been a major topic of discussion \cite{rokaj2018light,galego2019cavity,schafer2020relevance,fregoni2022theoretical,delapradilla2025there}. The current consensus is that generally, for Fabry--P\'erot cavities it should not be discarded; in \emph{ab initio} calculations, its presence is required for there to be a stable electronic ground state \cite{rokaj2018light,schafer2020relevance}. In the context of thermal rate calculations, the DSE term plays an important role: it (nearly) exactly compensates for the energy shift arising from the linear light--matter coupling term \cite{li2020origin,yang2021quantum,mandal2023microscopic,fiechter2024understanding}, so that barrier heights are hardly affected by the cavity. 

Using that $\hat{\mu} = \hat{\mu}_A + \hat{\mu}_B$ for two molecules $A$ and $B$, it is clear that the DSE term will feature cross terms proportional to $\hat{\mu}_A \hat{\mu}_B$. A striking feature of these cross terms is that they seem to facilitate an instantaneous, distance-independent coupling between all molecules in the cavity \cite{schnappinger2023PES,haugland2023understanding,fischer2024cavity}. This may in turn have important consequences for the dynamics \cite{sidler2024unraveling,borges2024extending}. However, it is often overlooked that in the ``standard" Hamiltonian used in the field, intermolecular Coulombic interactions are neglected, even though it is well-known that in free space and if the molecules are far enough apart, these \emph{cancel} the intermolecular DSE cross terms \cite{cohen1997photons,keelingQO}. This cancellation has also been shown to hold for the case of a perfect cavity  \cite{power1982quantum,vukics2012adequacy,de2018cavity}, although these derivations may not seem particularly intuitive. What all these derivations have in common though, is that they go beyond the single-mode approximation, and take all light modes into account; this is a crucial ingredient to the cancellation. 
In any case, this cancellation calls into question if and how these seemingly important DSE cross terms should be included in a (single-mode) Hamiltonian when studying polaritonic chemistry.

In this paper, we revisit how this cancellation of intermolecular Coulombic interactions and DSE cross terms comes about in free space, and in a perfect cavity (see also Fig.~\ref{fig:summary}) -- we remark here already that in any event, the DSE self terms remain, and that what counts as a self term or a cross term depends on how far apart the molecules are. 
We then move on to study the argument in the case of an imperfect cavity, where we present numerical evidence suggesting that the cancellation may extend to this case as well. 
Finally, if one wants to work with realistic mirror materials, it turns out to be most practical to work in the framework of macroscopic quantum electrodynamics (macroscopic QED) \cite{buhmann2013dispersion_book}. This theory can correctly account for the absorption of light by the mirror material, and for the fluctuations that accompany this dissipation. As before, the direct (free-space-like) interactions between the molecules cancel, but it turns out that the cavity-mediated interactions are now affected by the aforementioned fluctuations; they are more naturally absorbed in the matter-field interaction term, and attempting to extract and cancel them can only complicate the description. Finally, we discuss the implications of these findings for the Hamiltonian used in studies on polaritonic chemistry.
With this, we hope to give more insight into the components of the Hamiltonian and the underlying approximations, and provide a framework for future studies.

\section{Cancellation in free space} \label{sec:freespace}
We begin our discussion by showing explicitly how Coulombic dipole--dipole interactions between two neutral, well-separated molecules exactly cancel out the dipole self-energy cross terms in free space, partly following the derivation in Ref. \cite{keelingQO}.
We work in atomic units, \textit{i.e.} $\hbar=e=4\pi \varepsilon_0 = 1$. 

We start out in the Coulomb gauge with the Pauli--Fierz Hamiltonian (excluding magnetic terms) \cite{cohen1997photons,ruggenthaler2023understanding}, 
\begin{equation}
    \hat{H}_\mathrm{PF}=\sum_\alpha\frac{1}{2m_\alpha} \Big[ \hat{\mathbf{p}}_\alpha {- q_\alpha\hat{\mathbf{A}}_\perp(\hat{\mathbf{r}}_\alpha) }\Big]^2 +\hat{V}_\mathrm{Coulomb} + \hat{H}_\mathrm{ph},
\end{equation}
where the sum is over particles with bare mass $m_\alpha$ and charge $q_\alpha$. In the Coulomb gauge, the vector potential is purely transverse; in free space, it is given by 
\begin{equation}
    \hat{\mathbf{A}}_{\perp}^{\mathrm{free}} (\mathbf{r}) 
    = \sum_{\mathbf{k},\lambda} {\mathbf{e}}_\lambda A_\mathbf{k} \Big[ e^{i \mathbf{k}\cdot \mathbf{r}} \hat{a}_{\mathbf{k},\lambda} + e^{-i \mathbf{k}\cdot \mathbf{r}} \hat{a}^\dagger_{\mathbf{k},\lambda}  \Big],
\end{equation}
where $\hat{a}_{\mathbf{k},\lambda}^\dagger$ ($\hat{a}_{\mathbf{k},\lambda}$) creates (annihilates) a photon with wavevector $\mathbf{k}$ and transverse polarization ${\mathbf{e}}_{\mathbf{k},\lambda}$, and $A_\mathbf{k}=\sqrt{{2\pi}/{\mathcal{V} \omega_\mathbf{k}}}$ with $\mathcal{V}$ the quantization volume. The energy of the free electromagnetic field is described by $\hat{H}_\mathrm{ph}={\sum_{\mathbf{k},\lambda} \omega_\mathbf{k} (\hat{a}^\dagger_{\mathbf{k},\lambda}\hat{a}_{\mathbf{k},\lambda}+\tfrac{1}{2})}$, where the sum goes over all wavevectors and both polarizations, and the Coulomb interaction between the charged particles is given by
\begin{equation}\label{eq:Vcoulomb}
\hat{V}_\mathrm{Coulomb} = \frac{1}{2}\sum_{\alpha\neq \beta} \frac{q_\alpha q_\beta}{|\hat{\mathbf{r}}_\alpha-\hat{\mathbf{r}}_\beta|}   . 
\end{equation}

We now specify our discussion to the case of two neutral molecules (localized clusters of charges), labeled $A$ and $B$. We will take both molecules to have a dipole moment, expressed as $\hat{\boldsymbol{\mu}}_i =\sum_{\alpha\in i} q_\alpha (\hat{\mathbf{r}}_\alpha-\mathbf{r}_i)$ for $i=A,B$, with $\mathbf{r}_i$ the center of mass of molecule $i$; note that the sum here runs over all electrons and nuclei that make up molecule $i$. The distance between $A$ and $B$ we set to be much larger than the size of either molecule (see Sec.~\ref{sec:LWA}). This will allows us to describe their interaction via their dipole moments only. Another implication of this is that in a chemical setting (in solution), we can consider $A$ and $B$ to be distinguishable and treat their center of mass coordinates classically.

Next, we transform to the multi-center dipole gauge with the transformation \footnote{Note that by doing this multi-center dipole transformation, we break the symmetry of the Hamiltonian under exchange of particles (second line of Eq.~\eqref{eq:fullHfree}): we effectively distinguish the charged particles (\emph{i.e.} electrons/nuclei) that make up dipole $\hat{\boldsymbol{\mu}}_A$ from those in dipole $\hat{\boldsymbol{\mu}}_B$ by assigning the former to be close to $\mathbf{r}_A$, and the latter to $\mathbf{r}_B$. In the context of electronic structure calculations, this means that we neglect the exchange interaction between electrons in different molecules; as this interaction decays exponentially, it is expected to be negligible if $A$ and $B$ are separated by a distance multiple times their size \cite{stone2013theory}, as assumed in the text. } 
\begin{equation} \label{eq:gaugeU-free}
\hat{U}_\mathrm{ED} = \exp [-i \hat{\boldsymbol{\mu}}_A \cdot  \hat{\mathbf{A}}_\perp(\mathbf{r}_A) - i \hat{\boldsymbol{\mu}}_B \cdot  \hat{\mathbf{A}}_\perp(\mathbf{r}_B) ],
\end{equation}
where we rely on the long-wavelength approximation (LWA), \textit{i.e.} we restrict ourselves to wavelengths much larger than the extent of either molecule. At this point, the question may arise when the two molecules can or should be regarded as separate entities in the LWA; we discuss this in more detail in Sec.~\ref{sec:LWA}.

On top of the gauge transformation, and in accordance with previous work, we perform a unitary rotation of the phase of the photonic modes by $\pi/2$, given by $\hat{U}_{\pi/2}=\exp{-i\tfrac{\pi}{2}\sum_{\mathbf{k},\lambda}\hat{a}_{\mathbf{k},\lambda}^\dagger\hat{a}_{\mathbf{k},\lambda}}$, which has the effect $\hat{a}_{\mathbf{k},\lambda}\rightarrow i \hat{a}_{\mathbf{k},\lambda}$, $\hat{a}_{\mathbf{k},\lambda}^\dagger\rightarrow -i \hat{a}^\dagger_{\mathbf{k},\lambda}$. Overall this can be summarized as \cite{keelingQO,mandal2023theoretical}
\begin{equation} \label{eq:UHU}
    \hat{H}^\mathrm{free}_\mathrm{ED} =  \hat{U}_{\pi/2}\hat{U}_\mathrm{ED} \hat{H}_\mathrm{PF}^\mathrm{free} \hat{U}^\dagger_\mathrm{ED}\hat{U}^\dagger_{\pi/2}
\end{equation}
which yields
\begin{equation}
\begin{aligned}\label{eq:fullHfree}
    \hat{H}^\mathrm{free}_\mathrm{ED} &=  \hat{H}_\mathrm{ph} + \sum_{i=A,B}\bigg[\sum_{\alpha \in i} \frac{\hat{\mathbf{p}}_\alpha^2}{2m_\alpha} + \hat{V}_{\mathrm{Coulomb},i}   \\
    &+ \sum_{\mathbf{k}, \lambda} {\mathbf{e}}_{\mathbf{k},\lambda}\cdot \hat{\boldsymbol{\mu}}_i \sqrt{\frac{2\pi\omega_\mathbf{k}}{\mathcal{V}}} \bigg( e^{i \mathbf{k} \cdot \mathbf{r}_i} \hat{a}_{\mathbf{k},\lambda} + e^{-i \mathbf{k} \cdot \mathbf{r}_i}\hat{a}_{\mathbf{k},\lambda}^\dagger\bigg) \\
    & + \sum_{\mathbf{k},\lambda}\frac{2\pi}{\mathcal{V}}({\mathbf{e}}_{\mathbf{k},\lambda}\cdot \hat{\boldsymbol{\mu}}_i)^2 \Big] \\
    & + \hat{V}^\mathrm{free}_{\mathrm{DSE}, AB} + \hat{V}^\mathrm{free}_{\text{Coulomb},AB}.
\end{aligned}
\end{equation}
Here, the second line describes the coupling of each of the dipoles to the transverse displacement (rather than electric) field, the third line contains the dipole self-energy term of each of the dipoles individually, and the last line describes the direct interactions between the two dipoles. The dipole self-energy cross term is given by
\begin{equation}
\begin{aligned} \label{eq:VDSEfree}
    \hat{V}^\mathrm{free}_{\mathrm{DSE}, AB}&=  \sum_{\mathbf{k}, \lambda}\frac{4\pi}{\mathcal{V}}(\mathbf{e}_{\mathbf{k},\lambda} \cdot \hat{\boldsymbol{\mu}}_A)(\mathbf{e}_{\mathbf{k},\lambda} \cdot \hat{\boldsymbol{\mu}}_B)e^{-i \mathbf{k} \cdot (\mathbf{r}_A-\mathbf{r}_B)} \\
    &= 2\sum_\lambda \int\frac{\dd^3\mathbf{k}}{(2\pi)^2}\, (\mathbf{e}_{\mathbf{k},\lambda}\cdot \hat{\boldsymbol{\mu}}_A) (\mathbf{e}_{\mathbf{k},\lambda}\cdot \hat{\boldsymbol{\mu}}_B)    e^{-i\mathbf{k}\cdot (\mathbf{r}_A - \mathbf{r}_B )}. \\
\end{aligned}
\end{equation}

We now turn to the intermolecular Coulomb interaction, given by \cite{cohen1997photons}
\begin{subequations}
\begin{align}
\hat{V}^\mathrm{free}_{\mathrm{Coulomb},AB} \label{eq:coulombABfree}&= \sum_{\alpha\in A,\, \beta \in B} \frac{q_\alpha q_\beta}{|\hat{\mathbf{r}}_\alpha-\hat{\mathbf{r}}_\beta|} \\
&= \iint \dd \mathbf{r} \, \dd \mathbf{r}' \frac{\hat{\rho}_A(\mathbf{r})\hat{\rho}_{B}(\mathbf{r}')}{|\mathbf{r} - \mathbf{r}'|}\\
 &=2\pi \int \dd^3 \mathbf{k} \frac{1}{k^2} \big(\hat{\rho}_A(\mathbf{k}) \hat{\rho}^*_B(\mathbf{k}) + \mathrm{c.c.} \big). \label{eq:Vcoul_kspace}
\end{align}
\end{subequations}
In the second line, we rewrote the Coulomb potential in terms the charge densities {of the molecules, given by} $\hat{\rho}_i(\mathbf{r})=\sum_{\alpha\in i} q_\alpha \delta(\mathbf{r}-\hat{\mathbf{r}}_\alpha)$ {for  $i=A,B$}; in the third line we used the Parseval--Plancherel identity {and the convolution theorem} to convert it to an integral over $\mathbf{k}$-space in terms of the Fourier transform of the charge density, 
\begin{equation} \label{eq:rhoFT}
    \hat{\rho}_i(\mathbf{k}) = \frac{1}{(2\pi)^{3/2}} \sum_{\alpha\in i} q_\alpha e^{-i\mathbf{k}\cdot\hat{\mathbf{r}}_\alpha}. 
\end{equation}

As we have assumed molecules $A$ and $B$ to be sufficiently far apart, we can approximate their full Coulombic interaction by the first term in the multipole series only, the dipole--dipole interaction: $\hat{V}_{\mathrm{Coulomb},AB}\approx \hat{V}_{\mathrm{dip-dip},AB}$. 
{By doing this, we do restrict ourselves to only part of the full Hilbert space on which the Hamiltonian (Eq.~\eqref{eq:fullHfree}) acts: this truncation of the multipole expansion is only valid for quantum states where the electrons and nuclei are indeed localized close to either $\mathbf{r}_A$ or $\mathbf{r}_B$ (\emph{i.e.} conform to the conditions outlined in Sec.~\ref{sec:LWA}). This means that formally, we cannot simply replace $\hat{V}_{\mathrm{Coulomb},AB}$ in Eq.~\eqref{eq:fullHfree} with $\hat{V}_{\mathrm{dip-dip},AB}$ \cite{schafer2020relevance}. It also implies that in the following, when adding and comparing operators, we have to keep in mind that these only act on this subset of states.}

To obtain an expression for the Coulombic dipole--dipole interaction, we replace the molecular charge densities in Eq.~\eqref{eq:Vcoul_kspace} by {charge densities representing} their dipole moments only \footnote{Or more precisely, dipole moment matrix elements -- 
\emph{i.e.}, transition dipole moments as well as permanent dipole moments.}. {To construct such a charge density, we form a dipole with dipole moment $\boldsymbol{\mu}_i$ by placing} a positive charge of $+Q$ at position $\mathbf{r}_i + \tfrac{\boldsymbol{\mu}_i}{2Q}$ and a negative charge $-Q$ at  position $\mathbf{r}_i - \tfrac{\boldsymbol{\mu}_i}{2Q}$ \footnote{By increasing the value of $Q$, we can decrease the spatial extent of the dipoles, and eventually turn them into point dipoles; this guarantees us that their interaction is purely a dipole--dipole interaction.}.
{Following Eq.~\eqref{eq:rhoFT}}, the dipolar part of the molecular charge density can be expressed as
\begin{equation} \label{eq:rho_dip}
\begin{aligned}
    \hat{\rho}^\mathrm{dip}_i(\mathbf{k})
    &= \frac{Q}{(2\pi)^{3/2}} e^{-i\mathbf{k}\cdot \mathbf{r}_i} (e^{-i\mathbf{k}\cdot \hat{\boldsymbol{\mu}}_i /2Q} - e^{i\mathbf{k}\cdot \hat{\boldsymbol{\mu}}_i /2Q}) \\
    &\approx  \frac{-i}{(2\pi)^{3/2}} e^{-i\mathbf{k}\cdot \mathbf{r}_i} (\mathbf{k}\cdot \hat{\boldsymbol{\mu}}_i),
\end{aligned}
\end{equation}
as in the long-wavelength approximation $\mathbf{k}\cdot \boldsymbol{\mu}_i \ll 1$, which we can always satisfy by choosing $Q$ large enough. 

Substituting Eq.~\eqref{eq:rho_dip} into Eq.~\eqref{eq:Vcoul_kspace}, we obtain the Coulombic dipole--dipole interaction between $A$ and $B$,
\begin{equation} \label{eq:Vdipdip_kspace}
\hat{V}^\mathrm{free}_{\mathrm{dip-dip},AB}=2\int \frac{\dd^3\mathbf{k}}{(2\pi)^2}\, (\boldsymbol{\kappa}\cdot \hat{\boldsymbol{\mu}}_A) (\boldsymbol{\kappa}\cdot \hat{\boldsymbol{\mu}}_B)    e^{-i\mathbf{k}\cdot (\mathbf{r}_A - \mathbf{r}_B )},
\end{equation}
where $\boldsymbol{\kappa}=\mathbf{k}/k=(\sin\theta\cos\phi,\sin\theta\sin\phi,\cos\theta)$. 
Note that Eq. \ref{eq:Vdipdip_kspace} is equal to the more familiar real-space expression $\hat{V}^\mathrm{free}_{\mathrm{dip-dip},AB} = r^{-3}(\hat{\boldsymbol{\mu}}_A\cdot \hat{\boldsymbol{\mu}}_B - 3(\hat{r}\cdot\hat{\boldsymbol{\mu}}_A)(\hat{r}\cdot\hat{\boldsymbol{\mu}}_B))$, where $\hat{r}$ is the unit vector along $\mathbf{r}_A-\mathbf{r}_B$ and $r$ its magnitude; this is shown explicitly in the Supplementary Material.

The attentive reader will have noted the striking similarity between $\hat{V}^\mathrm{free}_{\mathrm{DSE}, AB}$ (Eq.~\eqref{eq:VDSEfree}) and $\hat{V}^\mathrm{free}_{\mathrm{dip-dip},AB}$ (Eq.~\eqref{eq:Vdipdip_kspace}). In fact, they can be shown to cancel exactly \cite{keelingQO,cohen1997photons}. As $\boldsymbol{\kappa}$ and the two $\mathbf{e}_{\mathbf{k},\lambda}$ together span the space, it holds that $(\boldsymbol{\kappa}\cdot \hat{\boldsymbol{\mu}}_A) (\boldsymbol{\kappa}\cdot \hat{\boldsymbol{\mu}}_B) + \sum_\lambda(\mathbf{e}_{\mathbf{k},\lambda}\cdot \hat{\boldsymbol{\mu}}_A) (\mathbf{e}_{\mathbf{k},\lambda}\cdot \hat{\boldsymbol{\mu}}_B) = \hat{\boldsymbol{\mu}}_A \cdot \hat{\boldsymbol{\mu}}_B $ for all $\mathbf{k}$, so that
\begin{equation} \label{eq:DSEdipdip_free}
\begin{aligned}
   \hat{V}^\mathrm{free}_{\mathrm{DSE}, AB} + \hat{V}^\mathrm{free}_{\mathrm{dip-dip}, AB} &= 2 \hat{\boldsymbol{\mu}}_A \cdot \hat{\boldsymbol{\mu}}_B \int\frac{\dd^3\mathbf{k}}{(2\pi)^2}\,e^{-i\mathbf{k}\cdot (\mathbf{r}_A - \mathbf{r}_B )}\\
   & \propto \delta(\mathbf{r}_A - \mathbf{r}_B ) = 0 
\end{aligned}
\end{equation}
where the delta function evaluates to 0 as long as our molecules $A$ and $B$ are well-separated, as discussed in more detail in Sec.~\ref{sec:LWA}.  An interpretation of this is that in free space in the electric-dipole gauge, there are \textit{no direct interactions}; all interactions are mediated by the fields (third line in Eq.~\eqref{eq:fullHfree}) \cite{keelingQO, cohen1997photons}.

Finally, at this point, we can already remark that in free space,
the DSE cross terms do \textit{not} represent distance-independent interactions; instead, from its relation to $V^\mathrm{free}_{\mathrm{dip-dip}, AB}$ we deduce that $V^\mathrm{free}_{\mathrm{DSE},AB} \propto r^{-3}$. In fact, we could have already come to this conclusion by performing the integral in Eq.~\eqref{eq:VDSEfree} analytically, as demonstrated in the Supplementary Material. Apparent distance-independence would arise from taking the single-mode approximation, where one effectively only retains a single term in the Fourier series and thereby loses real-space information. In Sec.~\ref{sec:perfcav}, we will see that this reasoning extends to the case of a perfect cavity. 

\subsection{Conditions on the separation between the molecules}\label{sec:LWA}
The dipole self-energy cross term that we are interested in only arises because we treated the two molecules as two separate entities: in Eq.~\eqref{eq:gaugeU-free}, we distinguished between the field at $\mathbf{r}_A$ and that at $\mathbf{r}_B$. This is necessary as soon as the distance between molecules $A$ and $B$ approaches a fraction of the wavelength of interest. When summing over all modes, this ``wavelength of interest" is set by the shortest wavelength/largest value of $k$ that we take into account: for the LWA to remain valid for each molecule individually, we should cut off the sums and integrals at some $k_\mathrm{max}$ such that $1/k_\mathrm{max}$ is still large compared to one molecule \cite{cohen1997photons}. Altogether this means that we should consider the molecules as separate entities in this context \emph{when they are separated by a distance much larger than their size}. 

Note that this coincides with the requirement for the full Coulombic interaction to be well-approximated with the dipole--dipole terms only, which is an essential ingredient in the argument presented above. Moreover, we remark that a cutoff on the integral in Eq.~\eqref{eq:DSEdipdip_free} will lead to a smearing out of the delta function over this $1/k_\mathrm{max}$ length scale \cite{cohen1997photons}. This means that it will only evaluate to zero if the molecules are well-separated in the sense outlined in the previous paragraph. 

{
In the previous two paragraphs, we have considered the ``size of a molecule", which one might have thought of as given by the positions of the nuclei at their equilibrium geometry in the electronic ground state (from now on referred to as ``classical size"). In quantum mechanics however, this ``size" in general depends on the electronic, vibrational and rotational state the molecule is in. The separation criterion thus becomes a statement about for which states the LWA/truncation of the multipole expansion is a good approximation: the extent of the electronic and nuclear wavefunction of either molecule should be much smaller than the distance between their centers of mass. This means that typically, low-lying vibrational or electronic excitations can be included in these states, but the same may not be true for \emph{e.g.} Rydberg excitations or scattering states. We have also considered the center of mass of our molecules to be at $\mathbf{r}_A$ and $\mathbf{r}_B$; quantum-mechanically, this can be accomplished by only selecting states for which the center-of-mass expectation values for molecules $A$ and $B$ match $\mathbf{r}_A$ and $\mathbf{r}_B$. By doing this, we essentially eliminate translations and rotations of the entire system.

It is important to recall that we have restricted our analysis of the cancellation to this set of states. For these states, we have shown that the energetic contributions of the Coulombic and DSE cross terms cancel, so that when acting with $\hat{H}_\mathrm{ED}^\mathrm{free}$ on one of these states, one could just as well have left both operators $\hat{V}^\mathrm{free}_{\mathrm{DSE}, AB}$ and $\hat{V}^\mathrm{free}_{\text{Coulomb},AB}$ out of the equation from the start. The question remains whether this set of states includes the states of interest, \emph{i.e.} the low-lying eigenstates or resonances of Eq.~\eqref{eq:fullHfree} (assuming we have found a way to do the renormalization that is required to obtain these) in the case that $A$ and $B$ are well-separated according to their classical size. This does seem plausible; after all, we know that these states should be similar to the low-lying rotational, vibrational or electronic excitations that we are familiar with from standard quantum chemistry, and typically for such states, the extent of both the nuclear and electronic wavefunctions is not much larger than their classical size. }

One could perform a consistency check here by finding the ground state in this restricted state space: it needs to obey the physical requirement that the expectation value of the transverse electric field vanishes \cite{schafer2020relevance}. If it violates this requirement, it means that true ground state cannot be fully represented in the restricted state space, so that the cancellation argument breaks down.

For completeness, we note that if one is willing to work with polarization densities, one can avoid taking the LWA and essentially show that all direct intermolecular interactions vanish up to all orders in multipole expansion (\emph{cf.}~Sec.~\ref{sec:mQED}), as long as the charge densities do not overlap \cite{cohen1997photons,salam2009molecular}. {However, this framework does come with some mathematical pitfalls \cite{woolley2024infinities}. }

\section{Cancellation in a perfect cavity}  \label{sec:perfcav}

We now move on to study the intermolecular Coulombic and dipole self-energy interactions between two molecules in a perfect cavity of length $L$. The cavity will introduce additional contributions both to the longitudinal (Coulombic) and transverse interactions. As mentioned, it has been shown before that also these additional contributions cancel out exactly \cite{power1982quantum, vukics2012adequacy}. Our main aim in this section is to express both terms in a formulation similar to the free-space discussion above, so that it becomes more obvious how exactly this cancellation comes about. 

\begin{figure}
    \centering
    \includegraphics[width=1.0\linewidth]{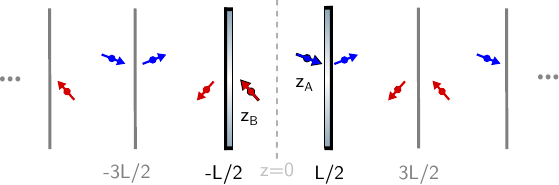} %
        \caption{Schematic of the image dipoles used to calculate the cavity-mediated part of the electrostatic interactions between molecules $A$ and $B$ using Eq.~\eqref{eq:images}.}
    \label{fig:images}
\end{figure}

We first study how the presence of mirrors modifies the Coulomb interaction. We begin by realizing that a charge in the proximity of a medium will induce some charge on the surface of the medium. The resulting field can be described by the method of images \cite{schwinger2019classical,einziger2002rigorous,Barcellona_2018} \footnote{The method of images is a well-known approach in classical electrodynamics, but can also be derived in the framework of quantum electrodynamics, where it is associated with the exchange of a virtual photon between a charge and an image charge \cite{Barcellona_2018}. Do note that this latter derivation makes use of perturbation theory. }  (Fig.~\ref{fig:images}). {As in the previous section, we will approximate the Coulomb interaction (now between a molecule and an image of the other molecule) by its dipole--dipole component only; for this to be valid, both molecules have to be far enough away from the mirrors.} The indirect interaction between molecules $A$ and $B$ via the mirrors is then given by
\begin{equation}\label{eq:images}
\begin{aligned}
    \hat{V}_{\mathrm{dip-dip}, AB}^{\mathrm{cav}} =& \tfrac{1}{2} (\hat{V}_{\mathrm{dip-dip}, A- \text{im}B} + \hat{V}_{\mathrm{dip-dip}, B- \text{im}A} ) \\
    =& \hat{V}_{\mathrm{dip-dip}, B- \text{im}A}
\end{aligned}
\end{equation}
where the factor of $\tfrac{1}{2}$ is introduced because for interactions between a real charge and an image charge, the potential energy is only half of the interaction energy between two real charges would be (see Ref. \cite{griffiths2023ED} for a discussion of this). The second equality follows as $V_{A- \text{im} B} =V_{B-\text{im} A} $  \cite{Barcellona_2018}.

As depicted in Fig.~\ref{fig:images}, the cavity produces an infinite number of images, which can be subdivided into two types: the image dipoles resulting from an even number of reflections point in the same direction as the original dipole, so that $\boldsymbol{\mu}_A^\mathrm{even} = \boldsymbol{\mu}_A$, while the image dipoles resulting from an odd number of reflections are flipped with respect to their $x$ and $y$ direction: $\boldsymbol{\mu}_A^\mathrm{odd} = \mathbf{R}_{xy} \boldsymbol{\mu}_A$ with $\mathbf{R}_{xy}=\mathrm{diag}(-1,-1,1)$.

Accounting for the position of the images, we obtain the charge density of the images of dipole $A$ as
\begin{equation}
\begin{aligned}\label{eq:rhoimA}
    \hat{\rho}_{\mathrm{im}A} =& \frac{-i}{(2\pi)^{3/2}} e^{-i\mathbf{k}_\parallel \cdot \mathbf{r}_A} \\&\bigg\{
    e^{-ik_z z_A}(\mathbf{k}\cdot \hat{\boldsymbol{\mu}}_A)\Big[\sum_{n=-\infty}^{\infty} e^{ik_z2nL} - 1\Big] \\
    &
    +e^{ik_z z_A}(\mathbf{k}\cdot \mathbf{R}_{xy} \hat{\boldsymbol{\mu}}_A)\sum_{n=-\infty}^\infty e^{ik_zL(2n-1)}  \bigg\}
\end{aligned}
\end{equation}
Here, $\mathbf{k}_\parallel = (k_x, k_y, 0)$ is the component of the wavevector parallel to the $xy$-plane (the plane of the mirrors), and $k_z$ is the component perpendicular to it. 

We can now substitute this charge density into Eq. \ref{eq:Vcoul_kspace} to calculate the interaction between dipole $B$ and the images of dipole $A$. We can perform the $k_z$ integral by recognizing that the infinite sums in Eq.~\eqref{eq:rhoimA} constitute Dirac combs, \emph{i.e.}
\begin{equation} \label{eq:diraccomb}
\frac{L}{\pi}\sum_{n=-\infty}^\infty e^{ik_z2nL} = \sum_{m=-\infty}^\infty \delta\Big(k_z - m\frac{\pi}{L}\Big).
\end{equation}

Adding the direct interaction (Eq.~\eqref{eq:Vdipdip_kspace}) to the cavity-mediated interaction, we obtain the total Coulombic dipole--dipole interaction in a cavity, 
\begin{widetext}
\begin{equation} \label{eq:Vcavlong}
\begin{aligned}
        \hat{V}^\mathrm{cav}_{\mathrm{dip-dip},AB}+\hat{V}^\mathrm{free}_{\mathrm{dip-dip},AB}
        =\frac{2\pi}{L}\int \frac{\dd^2\mathbf{k}_\parallel}{(2\pi)^2} \sum_{k_z}\bigg\{&(\boldsymbol{\kappa}\cdot \hat{\boldsymbol{\mu}}_A)(\boldsymbol{\kappa}\cdot \hat{\boldsymbol{\mu}}_B) e^{-i\mathbf{k}\cdot(\mathbf{r}_B-\mathbf{r}_A)} \\
        &+  (\boldsymbol{\kappa}\cdot\mathbf{R}_{xy} \hat{\boldsymbol{\mu}}_A) (\boldsymbol{\kappa} \cdot \hat{\boldsymbol{\mu}}_B) e^{-i\mathbf{k}_\parallel \cdot(\mathbf{r}_B-\mathbf{r}_A)}e^{i k_z(z_A+z_B+L)}   \bigg\},
    \end{aligned}
\end{equation}
\end{widetext}
where $k_z=m\pi/L$ with $m$ an integer running from $-\infty$ to $\infty$.

Next, we investigate the effect that the mirrors have on the transverse (DSE) interactions. The mirrors change the mode structure: instead of plane waves, the vector potential should now be expanded into the mode functions appropriate for a Fabry--P\'erot cavity. 
For a perfect cavity, these are given by \cite{meschede1990radiative, vukics2012adequacy}
\begin{subequations} 
\begin{align}
\label{eq:defmodefunc}
\Modefunc_{\mathbf{k}, \mathrm{s}}(z) &= \mathbf{e}_\mathrm{s} \sin(k_z [z+L/2]) \\ 
\begin{split}
\Modefunc_{\mathbf{k}, \mathrm{p}}(z) &= 
\tfrac{1}{2} \Big[(\mathbf{e}_\mathrm{p}  + \mathbf{R}_{xy}\mathbf{e}_\mathrm{p})\cos(k_z [z+L/2]) \\
& \hspace{0.4cm}+i (\mathbf{e}_\mathrm{p} - \mathbf{R}_{xy}\mathbf{e}_\mathrm{p})\sin(k_z [z+L/2])\Big] 
\end{split}
\end{align}
\end{subequations}
for s- and p-polarized light, respectively. Again, $k_z$ can only take values of $k_z=m\pi/L$, where $m$ is an integer. The transverse vector potential in the cavity can now be expressed as \footnote{Note that in contrast to Vukics \emph{et al.} \cite{vukics2012adequacy}, we let the sum over $k_z$ run over negative as well as positive values; our normalization $A_\mathbf{k}$ therefore deviates from theirs accordingly}
\begin{equation}
    \hat{\mathbf{A}}_{\perp}^{\mathrm{cav}}(\mathbf{r}) = \sum_{\mathbf{k},\lambda} A_\mathbf{k} \Big( e^{i\mathbf{k}_\parallel \cdot \mathbf{r}} \Modefunc_{\mathbf{k}, \lambda}(z) \hat{a}_{\mathbf{k}, \lambda} +  e^{-i\mathbf{k}_\parallel \cdot \mathbf{r}} \Modefunc^*_{\mathbf{k}, \lambda}(z) \hat{a}^\dagger_{\mathbf{k}, \lambda} \Big).
\end{equation}
This change in the vector potential will affect the gauge transformation in Eq. \ref{eq:gaugeU-free}, which now explicitly becomes 
\begin{equation}
    \hat{U}_\mathrm{ED}=\exp [-\sum_{\mathbf{k}, \lambda}i \Big(\hat{c}_{\mathbf{k}, \lambda}(\mathbf{r}_A, \mathbf{r}_B) \hat{a}_{\mathbf{k}, \lambda} + \hat{c}^*_{\mathbf{k}, \lambda}(\mathbf{r}_A, \mathbf{r}_B) \hat{a}^\dagger_{\mathbf{k}, \lambda} \Big)   ],
\end{equation}
where we defined 
\begin{equation} \label{eq:c}
\begin{aligned}
    \hat{c}_{\mathbf{k}, \lambda}(\mathbf{r}_A, \mathbf{r}_B) = A_\mathbf{k} (&e^{i\mathbf{k}_\parallel \cdot \mathbf{r}_A} \hat{\boldsymbol{\mu}}_A \cdot \Modefunc_{\mathbf{k}, \lambda}(z_A)  \\
    & + e^{i\mathbf{k}_\parallel \cdot \mathbf{r}_B} \hat{\boldsymbol{\mu}}_B \cdot \Modefunc_{\mathbf{k}, \lambda}(z_B)).
    \end{aligned}
\end{equation}
As before, this transformation commutes with the position operator, and therefore with $\hat{V}_\mathrm{Coulomb}$. Also the commutation with $\hat{\mathbf{p}}_\alpha$ is no different from the free-space case \cite{keelingQO}: in the long-wavelength approximation, 
$\hat{U}^\dagger \hat{\mathbf{p}}_\alpha \hat{U} \approx \hat{\mathbf{p}}_\alpha + q_\alpha \hat{\mathbf{A}}_{\perp, \mathrm{cav}} (\mathbf{r}_A)$
for $\alpha \in A$; for a particle in system $B$ one just replaces $\mathbf{r}_A$ by $\mathbf{r}_B$. The transformation of $\hat{H}_\mathrm{ph}$ will give us the linear light--matter interaction and DSE terms. Using the Baker--Campbell--Hausdorff theorem we find that
\begin{equation}
\begin{aligned}
    \hat{U}_\mathrm{ED}\hat{H}_\mathrm{ph} \hat{U}_\mathrm{ED}^\dagger =& \sum_{\mathbf{k},\lambda} 
    \omega_{\mathbf{k},\lambda} \Big[\hat{a}^\dagger_{\mathbf{k}, \lambda}\hat{a}_{\mathbf{k}, \lambda}+ \tfrac{1}{2} + |\hat{c}_{\mathbf{k}, \lambda}(\mathbf{r}_A, \mathbf{r}_B)|^2 \\
    &+ i\hat{c}^*_{\mathbf{k}, \lambda}(\mathbf{r}_A, \mathbf{r}_B) \hat{a}^\dagger_{\mathbf{k}, \lambda} -i\hat{c}_{\mathbf{k}, \lambda}(\mathbf{r}_A, \mathbf{r}_B)\hat{a}_{\mathbf{k}, \lambda}    \Big].
\end{aligned}
\end{equation}
The DSE interactions are described by the $\sum_{\mathbf{k},\lambda} 
\omega_{\mathbf{k},\lambda} |\hat{c}_{\mathbf{k}, \lambda}(\mathbf{r}_A, \mathbf{r}_B)|^2 $ term. After some algebra \footnote{We used here that we can flip the signs of $k_z$ and $\mathbf{k}_\parallel$ under the integral. Flipping the sign of $k_z$ has the effect $\mathbf{e}_\mathrm{p}\rightarrow \mathbf{R}_{xy}\mathbf{e}_\mathrm{p}$ while leaving $\mathbf{e}_\mathrm{s}$ unchanged. We have also used that $\mathbf{R}_{xy}\mathbf{e}_\mathrm{s}=-\mathbf{e}_\mathrm{s}$. These properties can easily be verified using the explicit form of the polarization vectors given in the SI}, we find that the $A$--$B$ interactions arising from this are given by 
\begin{widetext}
\begin{equation} \label{eq:Vcavtrans}
\begin{aligned}
        \hat{V}_{\mathrm{DSE},AB}=\hat{V}^\mathrm{cav}_{\mathrm{DSE},AB}+\hat{V}^\mathrm{free}_{\mathrm{DSE},AB}=\frac{2\pi}{L}\int \frac{\dd^2 \mathbf{k}_\parallel}{(2\pi)^2} \sum_{k_z} \sum_\lambda \Bigg\{&(\mathbf{e}_{\lambda}\cdot \hat{\boldsymbol{\mu}}_A)(\mathbf{e}_{\lambda}\cdot \hat{\boldsymbol{\mu}}_B)e^{-i\mathbf{k} \cdot (\mathbf{r}_B-\mathbf{r}_A)}  \\
        &+ (\mathbf{e}_{\lambda}\cdot\mathbf{R}_{xy} \hat{\boldsymbol{\mu}}_A)(\mathbf{e}_{\lambda}\cdot \hat{\boldsymbol{\mu}}_B)  e^{-i\mathbf{k}_\parallel \cdot (\mathbf{r}_B-\mathbf{r}_A)} e^{ik_z (z_A+z_B+L)} \Bigg\},
    \end{aligned}
\end{equation}
\end{widetext}
where for later convenience, we have defined the cavity contribution to the DSE cross term as $\hat{V}_{\mathrm{DSE},AB}^\mathrm{cav}=\hat{V}_{\mathrm{DSE},AB}-\hat{V}_{\mathrm{DSE},AB}^\mathrm{free}$, with $\hat{V}_{\mathrm{DSE},AB}^\mathrm{free}$ given by Eq.~\eqref{eq:VDSEfree}.

As in free-space, Eq. \ref{eq:Vcavtrans} bears a great similarity to Eq. \ref{eq:Vcavlong}, encouraging us  perform the same trick as before (Eq. \ref{eq:DSEdipdip_free}); indeed, one can show that also $(\boldsymbol{\kappa}\cdot\mathbf{R}_{xy} \hat{\boldsymbol{\mu}}_A) (\boldsymbol{\kappa}\cdot \hat{\boldsymbol{\mu}}_B) + \sum_\lambda(\mathbf{e}_{\mathbf{k},\lambda}\cdot\mathbf{R}_{xy} \hat{\boldsymbol{\mu}}_A) (\mathbf{e}_{\mathbf{k},\lambda}\cdot \hat{\boldsymbol{\mu}}_B) = \mathbf{R}_{xy}\hat{\boldsymbol{\mu}}_A \cdot \hat{\boldsymbol{\mu}}_B $, which is $\mathbf{k}$-independent and can therefore be pulled out of the integral. For the remaining integrals over rotating exponentials, the main difference to free space is the fact that we sum over discrete $k_z$, rather than integrate over continuous values. This gives rise to Dirac combs; \emph{cf.}~Eq.~\eqref{eq:diraccomb}, but now in real space. The first term is proportional to
\begin{equation}
\begin{aligned}
    \delta(x_A-x_B)\delta(y_A-y_B)\sum_n \delta(z_A-z_B- 2nL), \\
    \end{aligned}
\end{equation}
which clearly vanishes. This follows from our assumption that molecules $A$ and $B$ are well-separated, which means that the $n=0$ term in the sum vanishes, and from the fact that both molecules are in the cavity, which means that their $z$-coordinates cannot differ by more than $L$. The second term is proportional to 
\begin{equation}
\begin{aligned}
    \delta(x_A-x_B)\delta(y_A-y_B)\sum_n \delta(z_A+z_B+L- 2nL); \\
    \end{aligned}
\end{equation}
this vanishes too: as $-L/2<z_A<L/2$ and similarly for $z_B$, it is clear that $0<z_A+z_B+L<2L$, meaning the delta function will always return 0. With this, we have shown that the intermolecular Coulombic and DSE interactions still cancel in a perfect cavity. Moreover, as we know from  Sec.~\ref{sec:freespace} that the free-space contributions cancel individually, $\hat{V}^\mathrm{free}_{\mathrm{DSE},AB}+\hat{V}^\mathrm{free}_{\mathrm{dip-dip},AB}=0$; we can conclude that the cavity-mediated contributions separately vanish too, \begin{equation}
    \hat{V}^\mathrm{cav}_{\mathrm{DSE},AB}+\hat{V}^\mathrm{cav}_{\mathrm{dip-dip},AB}=0.
\end{equation}
{As discussed in Sec~\ref{sec:LWA}, one should keep in mind that this only holds for a restricted set of states for which we can regard the molecules as well-separated. Additionally, as mentioned before, each molecule should be far enough away from the mirrors, so that its Coulombic interaction with its images can be described via the dipole--dipole interaction only.}

From this cancellation it is clear that also in a cavity, the DSE cross terms do not represent distance-independent interactions. As discussed before, the free space contribution ${V}^\mathrm{free}_{\mathrm{dip-dip},AB}\propto r^{-3}$, meaning that also ${V}^\mathrm{free}_{\mathrm{DSE},AB}\propto r^{-3}$. Similarly, we know that the image interaction ${V}^\mathrm{cav}_{\mathrm{dip-dip},AB}$ depends on the relative positions of $A$ and $B$ and their distances to the mirrors, which means that ${V}^\mathrm{cav}_{\mathrm{DSE},AB}$ must do so too. 
As noted in the previous section, it is the single-mode approximation that removes spatial information and renders DSE cross terms seemingly independent of position. 

Before closing this section, we give the full Hamiltonian in electric-dipole gauge (after applying $\hat{U}_{\pi/2}$, \emph{cf.}~Eq.~\eqref{eq:UHU}) for the sake of completeness:
\begin{equation}\label{eq:fullHEDlike7}
\begin{aligned}
    \hat{H}_\mathrm{ED} &=  \hat{H}_\mathrm{ph} + \sum_{i=A,B}\bigg[\sum_{\alpha \in i} \frac{\hat{\mathbf{p}}_\alpha^2}{2m_\alpha} + \hat{V}^\mathrm{free}_{\mathrm{Coulomb},i} +\hat{V}^\mathrm{cav}_{\mathrm{Coulomb},i}  \\
    &+ \sum_{\mathbf{k}, \lambda} \sqrt{\frac{2\pi\omega_\mathbf{k}}{\mathcal{V}}} \Big( e^{i\mathbf{k}_\parallel \cdot \mathbf{r}_i} \hat{\boldsymbol{\mu}}_i \cdot \Modefunc_{\mathbf{k}, \lambda}(z_i) \hat{a}_{\mathbf{k}, \lambda}  + \mathrm{h.c.}\Big)\\
    & + \sum_{\mathbf{k},\lambda}\frac{2\pi}{\mathcal{V}}|\boldsymbol{\mu}_i \cdot \Modefunc_{\mathbf{k}, \lambda}(z_i)|^2 \Big]  + \hat{V}^\mathrm{free}_{\mathrm{DSE}, AB} + \hat{V}^\mathrm{free}_{\text{Coulomb},AB} \\
    & + \hat{V}^\mathrm{cav}_{\mathrm{DSE}, AB} + \hat{V}^\mathrm{cav}_{\text{Coulomb},AB}.
\end{aligned}
\end{equation}
where we have included all terms explicitly, also the terms that we have argued can cancel. This means that the expression is valid for the full Hilbert space (and not just the restricted set of states). Note that also the intramolecular Coulomb interaction is affected by the cavity: it is straightforward to construct $\hat{V}^\mathrm{cav}_{\mathrm{Coulomb},i}$ using the method of images. 

We can also make the interaction term linear in $\hat{q}_{\mathbf{k},\lambda}\propto \hat{a}_{\mathbf{k},\lambda}+a_{\mathbf{k},\lambda}^\dagger$ by performing a different unitary rotation $\hat{U}_{\phi-\pi/2}$ (given in the Supplemental Material). After converting the photonic creation and annihilation operators to canonical operators we obtain
\begin{equation}\label{eq:fullHED}
\begin{aligned}
    \hat{H}_\mathrm{ED} = & \sum_{i=A,B}\bigg[\sum_{\alpha \in i} \frac{\mathbf{p}_\alpha^2}{2m_\alpha} + \hat{V}_{\mathrm{Coulomb},i} \bigg] + \hat{V}_{\mathrm{Coulomb}, AB} \\
    &+\sum_{\mathbf{k}, \lambda} 
    \bigg[ \frac{\hat{p}^2_{\mathbf{k},\lambda}}{2} +\frac{1}{2} \omega_{\mathbf{k}}^2 \Big( \hat{q}_{\mathbf{k},\lambda} + \sqrt{\frac{2} 
    {\omega_{\mathbf{k}}}}\big| \hat{c}_{\mathbf{k}, \lambda}(\mathbf{r}_A, \mathbf{r}_B) \big| \Big)^2 \bigg] ,
\end{aligned}
\end{equation}
where for the sake of brevity we have now absorbed the free-space and cavity-mediated contributions to the Coulombic interactions in a single term $\hat{V}_\mathrm{Coulomb}$. It is straightforward to generalize this expression to the case of $N$ molecules (which enter via Eq.~\eqref{eq:c}). 

\section{Extension to imperfect cavities}
\subsection{Mode-function based description}
\begin{figure}
    \centering
    \includegraphics[width=1.0\linewidth]{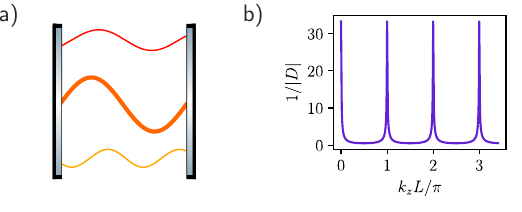}
    \caption{Mode functions for the vector potential in an imperfect cavity. a) As the transverse electric field is now allowed to be non-zero at the mirror surface, also modes that do not fit exactly between the mirrors (have a $k_z$ slightly different from $k_z=m\pi/L$) are allowed to exist, although they do have a smaller amplitude. b) The mode functions inherit the resonance behavior of $1/|D|$ (here plotted for a finesse of $\mathcal{F}\approx50$).  }
    \label{fig:imp_mode}
\end{figure}

In reality, no cavity is perfect: mirrors never reflect all of the incoming light. This means light in the cavity can leak out, which leads to spectral broadening of the cavity modes. In other words, modes with $k_z$ that nearly, but not exactly, fulfill the resonance condition $k_z=m\pi/L$ gain some amplitude in an imperfect cavity. This is encoded in the mode functions for the transverse field in an imperfect Fabry--Pérot cavity
\cite{ley1987quantum,de1991spontaneous,svendsen2023theory},
\begin{equation} \label{eq:modefunc_impf}
\Modefunc_{\mathbf{k}, \lambda}(z) = t_\lambda\frac{e^{ik_z z} \mathbf{e}_{\lambda} + r_\lambda e^{-ik_z z} e^{+i |k_z| L}\mathbf{e}^{-k_z}_{\lambda} }{\mathsf{D}_\lambda}  
\end{equation}
where $\lambda = \mathrm{s}, \mathrm{p}$ labels the polarization, and $\mathbf{e}^{-k_z}_{\lambda}$ is the polarization of a reflected wave traveling in the $-k_z$ direction. $t_\lambda$ and $r_\lambda$ are mirror transmission and reflection coefficients that are assumed to satisfy $|r_\lambda|^2+|t_\lambda|^2=1$ and $r_\lambda t_\lambda^*+r_\lambda^*t_\lambda=0$, and the denominator is given by $\mathsf{D}_\lambda = 1-r_\lambda^2\exp(2i|k_z|L)$. It is this denominator that is the origin of the large increase of the amplitude around a cavity resonance $k_z=m\pi/L$: if $r_\lambda^2$ is close to 1, $\mathsf{D}_\lambda$ becomes very small around resonant values of $k_z$, so that $\Modefunc_{\mathbf{k}, \lambda}(z)$ becomes very large in magnitude. This is illustrated in Fig.~\ref{fig:imp_mode}.

For dielectric mirrors, $t_\lambda$ and $r_\lambda$  can be expressed in terms of thickness of the mirror and its Fresnel coefficients; these Fresnel coefficients in turn depend on the angle of incidence and on the refractive index of the mirror material. Explicit relations for this are derived in Ref. \cite{dutra1996spontaneous}, and are also given in the Supplemental Material of this article. Given the reflection coefficients, we can calculate mirror reflectance via $R=|r(\theta=0,k=\pi/L)|^2$. The cavity's finesse can then be expressed as \cite{suter2014calculation}
\begin{equation}
    \mathcal{F} = - \frac{\pi}{\ln(R)}.
\end{equation}

The Hamiltonian for this case can be constructed by simply substituting the new mode functions (Eq.~\eqref{eq:modefunc_impf}) in the Hamiltonian given in Eq.~\eqref{eq:fullHEDlike7}, or in Eq.~\eqref{eq:fullHED} via the coupling operator $\hat{c}_{\mathbf{k},\lambda}$ (Eq.~\eqref{eq:c}). Additionally, one should make sure include the correct ``cavity-mediated" contributions to the Coulomb interactions, \emph{i.e.} the interactions between the molecules and the dielectric slabs that constitute the mirrors, as discussed below.

In order to investigate the cancellation, we focus on formulating explicit expressions for both the Coulombic dipole--dipole interaction and the DSE cross term. Obviously, the free-space parts of these interactions cancel as before; we can therefore focus on the cavity-mediated parts. 

It is straightforward to find an expression for the DSE: in Eq.~\eqref{eq:c}, one now just substitutes in the mode functions for an imperfect cavity (Eq.~\eqref{eq:modefunc_impf}). Then, as before, the DSE is given by $\sum_{\mathbf{k},\lambda} 
\omega_{\mathbf{k},\lambda} |\hat{c}_{\mathbf{k}, \lambda}(\mathbf{r}_A, \mathbf{r}_B)|^2 $; one can write this out and isolate the terms coupling $A$ and $B$, which represent the intermolecular DSE interactions. This results in an expression that bears great similarity to the case of the perfect cavity (Eq.~\eqref{eq:Vcavtrans}):
\begin{widetext}
\begin{equation} \label{eq:Vcavtrans_imp}
\begin{aligned}
       \hat{V}^\mathrm{imp.cav}_{\mathrm{DSE},AB}=2\int \frac{\dd^3 \mathbf{k}}{(2\pi)^2} \sum_\lambda \Bigg\{&(\mathbf{e}_{\lambda}\cdot \hat{\boldsymbol{\mu}}_A)(\mathbf{e}_{\lambda}\cdot \hat{\boldsymbol{\mu}}_B)e^{-i\mathbf{k} \cdot (\mathbf{r}_B-\mathbf{r}_A)} \bigg(\frac{|t_\lambda|^2(1+|r_\lambda|^2)}{|\mathsf{D}_\lambda(k_z)|^2}-1\bigg) \\
        &+ (\mathbf{e}_{\lambda}\cdot \hat{\boldsymbol{\mu}}_A)(\mathbf{e}_{\lambda}^{-k_z}\cdot \hat{\boldsymbol{\mu}}_B)  e^{-i\mathbf{k}_\parallel \cdot (\mathbf{r}_B-\mathbf{r}_A)} e^{ik_z (z_A+z_B)} \frac{2|t_\lambda|^2\mathrm{Re} (r_\lambda e^{+i|k_z|L})}{|\mathsf{D}_\lambda(k_z)|^2}  \Bigg\}.
    \end{aligned}
\end{equation}
\end{widetext}
(Be aware here that Eq.~\eqref{eq:Vcavtrans} contains both $\hat{V}^\mathrm{cav}_{\mathrm{DSE},AB}+\hat{V}^\mathrm{free}_{\mathrm{DSE},AB}$, whereas Eq.~\eqref{eq:Vcavtrans_imp} only represents the cavity-mediated part; the ``$-1$" in the first line is the result of subtracting the free-space part from the full DSE cross term.)

Next, we look at the intermolecular Coulombic interactions. The situation here is slightly more involved than in the case of a perfect cavity; as the mirrors have a finite width and a finite refractive index, the method of images breaks down. However, we can still solve the electrostatic problem by finding its static Green function, as outlined in for example Ref.~\cite{schwinger2019classical}. The result can be found in the Supplementary Material. 

If we follow the methodology we used before, we would now add these two terms and bring them into a single integral -- in free space and in a perfect cavity, we showed that this integral vanishes. Unfortunately, this merging of the terms into a single integral is far from trivial in the present case, because the expression for static Green function is very different in structure to the DSE interaction term. It therefore seems to us that this analytic approach is not feasible in this case; we leave a more detailed discussion for the Supplementary Material.

An alternative to this is numerical evaluation of both the static Green function and the DSE interaction for a range of specific configurations of dipoles $A$ and $B$ \footnote{By this we mean that we effectively calculate the expectation values of the DSE and Coulombic dipole--dipole interaction for a given state. For simplicity, we use a product state, $\ket{\psi}=\ket{\phi}_A\ket{\phi}_B$ for this, so that we can just replace the operators $\hat{\boldsymbol{\mu}}_i$ by their expectation value $\bra{\phi}_i \,\hat{\boldsymbol{\mu}}_i \ket{\phi}_i$. It is however straightforward to generalize this to a linear combination of states of molecules $A$ and $B$. } to see whether the contributions cancel. The static Green function is readily computed; on the other hand, the evaluation of the 3-dimensional integral in the expression for the DSE interaction is a much more daunting task. This is because the integrand is oscillatory and has many sharp features (peaks for each resonant value of $k_z$), and because the upper bound of integration has to be taken to infinity. Our strategy for dealing with this is outlined in the Supplementary Material. While this does allow us to evaluate the p-polarization part of the integral in Eq.~\eqref{eq:Vcavtrans_imp}, the s-polarization term remains infeasible to compute for numerical reasons; this is also elaborated on in the Supplementary Material. Because of this, we will restrict our numerical analysis to cases in which the two dipoles only couple via the p-polarization, \emph{i.e.} when at least one of the dipoles is aligned along the cavity axis $\hat{z}$ (note that $\mathbf{e}_\mathrm{s} \cdot \hat{z} = 0$). Whether the interactions cancel in general remains an open question. 

\begin{figure}
    \centering
    \includegraphics[width=\linewidth]{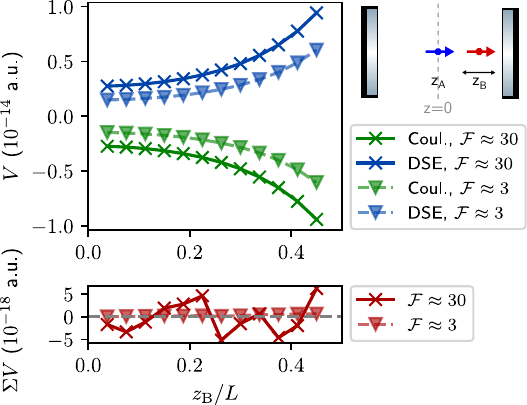}
    \caption{Cancellation of the cavity-mediated DSE cross terms (blue) and Coulombic dipole--dipole interactions (green) in an imperfect cavity; their sum is shown in red in the lower panel. Note that this configuration is special in the sense that only the p-polarization term contributes to the total energy. We have been unable to converge the s-polarization term to the integral in Eq.~\eqref{eq:Vcavtrans_imp} that would contribute in a more general case.  }
    \label{fig:impcav} 
\end{figure}

In Fig.~\ref{fig:impcav} we present numerical results for the interactions between two dipoles that are aligned along the cavity axis; the position of dipole $A$ is fixed at the center of the cavity, while the position of dipole $B$ is varied between 0 and $L/2$.  The Coulombic dipole--dipole interaction attracts dipole $B$ to the mirror; this can be understood qualitatively as $B$ being attracted to an image dipole of $A$ (\emph{cf.}~Fig.~\ref{fig:images}). The DSE cross terms show the exact opposite behaviour, in such a way that they cancel the Coulombic dipole--dipole interaction. For finesses up to about $\mathcal{F}\sim 50$, we find that the cancellation is exact up to three significant digits; the remaining difference can be ascribed to the limits on the accuracy of our numerical integration procedure. For higher finesses, larger numerical errors start creeping into the evaluation of the DSE integral; we do however see no clear reason to suspect that the cancellation actually breaks in this regime, it just becomes impossible to accurately evaluate the integral numerically. Results for a range of cavity finesses can be found in the Supplementary Material.

\subsection{Macroscopic-QED based description} \label{sec:mQED}

In a way, the situation considered in this section up till now is still not very realistic: we have assumed the refractive index of the mirrors to be constant across all frequencies, leading to a monotonic increase of the reflectance with frequency. However, it is well-known that the reflectance of a material is more highly frequency-dependent; gold for instance, often used as a mirror coating in polaritonic chemistry experiments, is only a good reflector up to about 600 nm \cite{thorlabsgold} -- for shorter wavelengths, the reflectivity drops. 
There is one major obstacle to incorporating this in our theory: a reflectivity cutoff requires the refractive index to be frequency-dependent, and according to the Kramers--Kronig relations \cite{zangwill2013modern}, this will lead to the refractive index acquiring an imaginary part. This results in absorption, and thereby breaks the lossless property $|r_\lambda|^2+|t_\lambda|^2=1$ and thus makes the mode-function formulation we used inapplicable. This should not come as a surprise; to capture absorption (dissipation) and the corresponding fluctuations, it is clear that the light modes have to be coupled to some sort of reservoir to funnel the energy away to; these are not included in our current description. We therefore move on to look at the problem in the framework of macroscopic QED \cite{buhmann2013dispersion_book, scheel2009macroscopic,feist2020macroscopic,hsu2025chemistry}, a theory which has been specifically designed to correctly quantize the electromagnetic field in the presence of absorbing materials.

For a comprehensive introduction, we refer the reader to Refs.~\cite{buhmann2013dispersion_book, scheel2009macroscopic}, but to attain a sense of the idea behind macroscopic QED, it is instructive to look at the Huttner-Barnett model \cite{scheel2009macroscopic,huttner1992quantization,huttner1992dispersion}. In short, in this model all of the light modes are coupled to a set of harmonic oscillators that represent the dielectric material (the mirror, in our case). These oscillators are in turn each coupled to their own reservoir; this facilitates absorption, as mentioned above. The model can be diagonalized to yield new normal modes, described by bosonic operators $\hat{\mathbf{f}}_\sigma(\mathbf{r},\omega)$ for each point in space, each frequency, and the electric and magnetic contributions $\sigma=e,m$. These operators, which represent mixed excitations of the electromagnetic field and of the absorbing dielectric, form the basic building blocks of macroscopic QED. 

\subsubsection{Coulomb gauge}
Let us now move on to look at how light--matter interaction is described in macroscopic QED. As before, our starting point is the Hamiltonian in the Coulomb gauge, which looks very similar to before, 
\begin{equation} \label{eq:H_mc_mQED}
\begin{aligned} 
    \hat{H}
    =&\hat{H}_\mathrm{ph} + \sum_\alpha\frac{1}{2m_\alpha} \Big[ \hat{\mathbf{p}}_\alpha {- q_\alpha\hat{\mathbf{A}}(\hat{\mathbf{r}}_\alpha) }\Big]^2 \\ 
    &+\hat{V}_\mathrm{Coulomb} + \sum_{i=A,B} \int \dd \mathbf{r} \hat{\rho}_i(\mathbf{r}) \hat{\phi}(\mathbf{r}).
\end{aligned}
\end{equation}
Here, the energy of the electromagnetic field is now described in terms of our new dynamic variables $\hat{\mathbf{f}}_\sigma(\mathbf{r},\omega)$,
\begin{equation}
    \hat{H}_\mathrm{ph} = \sum_\sigma \int_0^\infty \dd{\omega} \int \dd^3 \mathbf{r} \, \omega \,\hat{\mathbf{f}}^\dagger_\sigma(\mathbf{r},\omega) \hat{\mathbf{f}}_\sigma(\mathbf{r},\omega).
\end{equation}
Terms involving the vector potential $\hat{\mathbf{A}}(\mathbf{r})$ capture free-space as well as cavity-mediated interactions, whereas the term involving the scalar potential $\hat{\phi}(\mathbf{r})$ represents only cavity-mediated interactions -- in the language of the previous section, this term describes the interactions of the dipoles with their images. All direct (free-space) electrostatic interactions are already captured by $\hat{V}_\mathrm{Coulomb}$. The main difference between the mode-function based formulations above and macroscopic QED is that $\hat{\mathbf{A}}(\mathbf{r})$ and $\hat{\phi}(\mathbf{r})$ are now expressed in terms of our new dynamical variables $\hat{\mathbf{f}}_\sigma(\mathbf{r},\omega)$. This is done via 
\begin{align}
    \hat{\mathbf{A}}(\mathbf{r}) &= \int_0^\infty \dd \omega \frac{1}{i\omega}\hat{\mathbf{E}}^\perp(\mathbf{r},\omega) + \mathrm{h.c.},\\
    -\nabla \hat{\phi}(\mathbf{r}) &= \int_0^\infty \dd \omega \, \hat{\mathbf{E}}^\parallel(\mathbf{r},\omega) + \mathrm{h.c.},
\end{align}
\emph{i.e.} in terms of perpendicular and longitudinal parts \footnote{The longitudinal and perpendicular parts of a vector field $\mathbf{v}(\mathbf{r})$ are most easily obtained in $\mathbf{k}$-space \cite{cohen1997photons}, from its Fourier transform $\mathbf{v}(\mathbf{k})$: $\mathbf{v}^\parallel(\mathbf{k})=\boldsymbol{\kappa}[\boldsymbol{\kappa}\cdot\mathbf{v}(\mathbf{k})]$, and $\mathbf{v}^\perp(\mathbf{k})= \mathbf{v}(\mathbf{k})- \mathbf{v}^\parallel(\mathbf{k})$.} of the frequency components of the electric field,
\begin{equation}
    \hat{\mathbf{E}}(\mathbf{r},\omega)  = \sum_{\sigma=e,m} \int \dd \mathbf{r}' \,\mathbf{G}_\sigma (\mathbf{r},\mathbf{r}',\omega) \cdot \hat{\mathbf{f}}_\sigma(\mathbf{r}',\omega).
\end{equation}
The functions $\mathbf{G}_\sigma (\mathbf{r},\mathbf{r}',\omega)$ in this expression are directly related \cite{buhmann2013dispersion_book} to the dyadic Green's function $\mathbf{G} (\mathbf{r},\mathbf{r}',\omega)$, known from classical electrodynamics. It is here that the details of the shape and structure of the dielectric materials (in our case the mirrors) are encoded into the theory. 
For future reference, we note that electric field itself can be expressed as 
\begin{equation}
    \hat{\mathbf{E}}(\mathbf{r}) = \int_0^\infty \dd \omega \, \hat{\mathbf{E}}(\mathbf{r},\omega) + \mathrm{h.c.}
\end{equation}

\subsubsection{Power--Zienau--Woolley transformation}
Next, we transform to the Power--Zienau--Woolley (PZW) gauge. As noted before, this gauge comes with a series of mathematical pitfalls, as discussed in \emph{e.g.} Ref.~\cite{woolley2024infinities} \footnote{For example, for a polarization density defined as in Eq.~\eqref{eq:poldens}, $\hat{U}_\mathrm{PZW}$ is actually not a unitary transformation. Also, in this formulation, infinities arise in the polarization self-energy term; this is ultimately due to taking the square of a delta function, which is ill-defined. This goes to show in this case it is important to realize that delta `functions' are actually distributions, and should be treated as such. }. We proceed nevertheless, but remark that in the following, results should be interpreted as valid in perturbation theory only; this also means that we can work with the physical mass, rather than the bare mass. 

In the previous sections, we made the long-wavelength approximation at the same time as we performed the gauge transformation. This allowed us to describe the total charge distribution of the molecules using only their dipole moments, a familiar quantity to any chemist. However, in macroscopic QED it is customary to work with the polarization density, 
\begin{equation} \label{eq:poldens}
\hat{\mathbf{P}}_i(\mathbf{r}) = \sum_{\alpha \in i} q_\alpha \bar{\mathbf{r}}_\alpha \int_0^1 \dd u \, \delta (\hat{\mathbf{r}} - \mathbf{r}_i - u  \bar{\mathbf{r}}_\alpha ),
\end{equation} 
where $\bar{\mathbf{r}}_\alpha$ is measured relative to $\mathbf{r}_i$, which is the center of mass of molecule $i$). This quantity is a full description of the charge distribution, and effectively contains information on all orders of the multipole expansion \cite{cohen1997photons}. We will stick to using this polarization density for now (\emph{i.e.} not take the long-wavelength approximation just yet), and perform a PZW transformation \cite{cohen1997photons,woolley2024infinities} 
(which is essentially a generalization of the electric-dipole gauge transformation we used in the previous sections), 
\begin{equation}
    \hat{U}_\mathrm{PZW} = \exp[-i\sum_{i}\int\dd \mathbf{r} \, \hat{\mathbf{P}}_i(\mathbf{r}) \cdot \hat{\mathbf{A}}(\mathbf{r})].
\end{equation}
As in the mode-function based formulation above, this transformation has the effect of transforming $ \hat{\mathbf{p}}_\alpha \rightarrow \hat{\mathbf{p}}_\alpha + q_\alpha \hat{\mathbf{A}} (\mathbf{r}_\alpha)$ (neglecting magnetic terms), 
so that the vector potential is eliminated from the kinetic term. The action of the PZW transformation on $\hat{H}_\mathrm{ph}$ again leads to the linear light--matter coupling term (second line) and the free-space ``polarization self-energy" (PSE, third line) \cite{buhmann2013dispersion_book}:
\begin{equation} \label{eq:PZW_Hph}
\begin{aligned}
    \hat{U}_\mathrm{PZW} \hat{H}_\mathrm{ph}\hat{U}^\dagger_\mathrm{PZW} =& \sum_\sigma \int_0^\infty \dd{\omega} \int \dd^3 \mathbf{r} \, \omega \,\hat{\mathbf{f}}^\dagger_\sigma(\mathbf{r},\omega) \hat{\mathbf{f}}_\sigma(\mathbf{r},\omega) \\
    & - \sum_i\int \dd \mathbf{r} \,\hat{\mathbf{P}}_i^\perp (\mathbf{r}) \cdot \hat{\mathbf{E}} (\mathbf{r}) \\
    &+ 2\pi \int \dd \mathbf{r} \,\Big[\sum_i\hat{\mathbf{P}}_i^\perp (\mathbf{r})\Big]^2
\end{aligned}
\end{equation}
where the $\perp$-projection of $\hat{\mathbf{P}}$ effectively arises because $\hat{\mathbf{A}}$ is proportional to the transverse component of $\hat{\mathbf{E}}$. Note that now, the $\hat{\mathbf{f}}_\sigma(\mathbf{r},\omega)$ operators have been transformed in such a way that in the new gauge $\hat{\mathbf{E}} (\mathbf{r})$ represents the displacement field in the cavity. 

In the long-wavelength approximation, this PSE 
corresponds exactly to the free-space contribution to the DSE that we derived in Sec.~\ref{sec:freespace} \footnote{This is straightforward to show using the Parseval--Plancherel identity, and the aforementioned projection of the perpendicular component of a vector field in $\mathbf{k}$-space \cite{cohen1997photons}.}:
\begin{equation}
\begin{aligned}
    \hat{V}_\mathrm{PSE}^\mathrm{free} =& 2\pi  \int \dd \mathbf{r} \,\Big[\sum_i\hat{\mathbf{P}}_i^\perp (\mathbf{r})\Big]^2 \\ \stackrel{
    \text{LWA}}{=}& \sum_{i,j,\lambda}\int\frac{\dd^3{k}}{(2\pi)^2}\, (\mathbf{e}_{\mathbf{k},\lambda}\cdot \hat{\boldsymbol{\mu}}_i) (\mathbf{e}_{\mathbf{k},\lambda}\cdot \hat{\boldsymbol{\mu}}_j)    e^{-i\mathbf{k}\cdot (\mathbf{r}_j - \mathbf{r}_i )};
    \end{aligned}
\end{equation}
this expression contains the self-terms ($i=j$) as well as the cross terms ($i\neq j$). The cavity-mediated contribution to the self-energy (contained in Eq.~\eqref{eq:Vcavtrans} in the case of a perfect cavity) is now hidden in the light--matter coupling term on the second line of Eq.~\eqref{eq:PZW_Hph}. 

We have now discussed how the terms in the first line of the Hamiltonian in Eq.~\eqref{eq:H_mc_mQED} transform under the PZW transformation. The terms in the second line of that equation are left invariant by the transformation, but it is customary to still rewrite them in a way that allows for some simplification later. The Coulomb interaction can be rewritten in terms of the longitudinal part of the polarization \cite{buhmann2013dispersion_book}. For clarity, we divide it up into inter- and intramolecular terms:
\begin{subequations} 
\begin{align}
\begin{split}\label{eq:VABmultipolar} 
    \hat{V}^\mathrm{free}_{\mathrm{Coulomb},AB} &= \sum_{\alpha \in A,\beta \in B}\frac{q_\alpha q_\beta}{|\hat{\mathbf{r}}_\alpha-\hat{\mathbf{r}}_\beta|} \\
    & = 2\cdot 2\pi \int \dd \mathbf{r} \,\hat{\mathbf{P}}_A^\parallel (\mathbf{r}) \cdot \hat{\mathbf{P}}_B^\parallel (\mathbf{r}),  \\
\end{split}&\\
\begin{split}
    \hat{V}^\mathrm{free}_{\mathrm{Coulomb},A} &= \frac{1}{2}\sum_{\alpha ,\beta\in A}\frac{q_\alpha q_\beta}{|\hat{\mathbf{r}}_\alpha-\hat{\mathbf{r}}_\beta|} \\
    & = 2\pi \int \dd \mathbf{r} \,\Big[\hat{\mathbf{P}}_A^\parallel (\mathbf{r}) \Big]^2. \\
   \end{split}
\end{align}
\end{subequations}
It should be obvious that again, in the long-wavelength approximation, Eq.~\eqref{eq:VABmultipolar} reduces to the familiar $1/r^3$ dipole--dipole interaction. Another important thing to note here is that in order to rewrite the intramolecular Coulomb interaction in this way, we had to include terms in the sum for which $\alpha=\beta$. These terms represent the electrostatic energy of every of the particle interacting with itself, which is obviously a divergent quantity. As this is rather impractical in numerical calculations, we will stick with the $\alpha \neq \beta$ form (Eq.~\eqref{eq:Vcoulomb})
in our final formulation.

The last term in the Hamiltonian, which accounts for the cavity-mediated electrostatic interactions, essentially describes the interaction of each molecule $i$ with the scalar potential $\hat{\phi}(\mathbf{r})$ that the mirrors generate in reaction to all of the molecules in the cavity. It can similarly be described in terms of the longitudinal polarization \cite{buhmann2013dispersion_book}, 
\begin{equation} \label{eq:VcavcoulombMQED}
\begin{aligned}
\hat{V}^\mathrm{cav}_{\mathrm{Coulomb},AB}=&
    \sum_{i} \int \dd \mathbf{r} \hat{\rho}_i(\mathbf{r}) \hat{\phi}(\mathbf{r}) \\
    =& - \sum_{i}\int \dd \mathbf{r} \, \hat{\mathbf{P}}_i^\parallel  (\mathbf{r}) \cdot \hat{\mathbf{E}} ^\parallel (\mathbf{r}).
    \end{aligned}
\end{equation}

\subsubsection{Interpretation}
Let us start out by investigating the free-space part of the intermolecular Coulombic and polarization self-energy and  interactions:
\begin{equation} 
\begin{aligned}
\hat{V}^\mathrm{free}_{\mathrm{PSE},AB}+\hat{V}^\mathrm{free}_{\mathrm{Coulomb},AB} =&  4\pi \int \dd \mathbf{r} \,\Big[\hat{\mathbf{P}}_A^\perp  \cdot \hat{\mathbf{P}}_B^\perp  + \hat{\mathbf{P}}_A^\parallel \cdot \hat{\mathbf{P}}_B^\parallel \Big] \\
=& 4\pi \int \dd \mathbf{r} \,\hat{\mathbf{P}}_A  \cdot \hat{\mathbf{P}}_B  = 0;
\end{aligned}
\end{equation}
\emph{i.e.} the interactions cancel as long as the charge distributions (electron densities) of molecules $A$ and $B$ do not overlap \cite{cohen1997photons}. This cancellation is analogous to our discussion in Sec~\ref{sec:freespace}, except for the fact that we did not take the long-wavelength approximation here. This result may therefore be regarded as more general; it shows that in addition to the dipole--dipole interaction, also all higher-order multipole interactions between $A$ and $B$ vanish. 

We remark that within the long-wavelength approximation, the result obviously reduces to Eq.~\eqref{eq:DSEdipdip_free}.

In the case of a perfect cavity, we saw that also the cavity-mediated DSE interactions cancel with the cavity-mediated electrostatic interactions. Within macroscopic QED, whether this holds is not at all clear anymore. A major difference with the case of a perfect cavity (and also with the mode-function based description of an imperfect cavity) is obvious from Eq.~\eqref{eq:VcavcoulombMQED}: the cavity-mediated electrostatic interaction now also contains the field operators $\hat{\mathbf{f}}_\sigma(\mathbf{r},\omega)$, whereas before it only dependended on molecular (dipole) operators (\emph{cf.}~Eq.~\eqref{eq:Vcavtrans}). The reason for this probably lies in the fluctuation-dissipation theorem: the material's absorption is necessarily accompanied by fluctuations, in this case of the material's polarization \cite{buhmann2013dispersion_book}. These fluctuations affect the instantaneous charge distribution in the material, and therefore the electrostatic interactions of a molecule with the material. The magnitude of these fluctuations depends on the state of the whole body-field system described by the  $\hat{\mathbf{f}}_\sigma(\mathbf{r},\omega)$ operators. 
Note that these operators can be eliminated from the description by treating the interaction in second-order perturbation theory \cite{Barcellona_2018}; the theory then reduces to familiar classical electrostatics, \emph{e.g.} to the method of images. 

In short, it is not obvious whether the cavity-mediated interactions can be canceled in the framework of macroscopic QED; moreover, one may even wonder whether there is a point in trying to do so. After all, when adding the cavity-mediated electrostatic interaction (Eq.~\eqref{eq:VcavcoulombMQED}) to the light--matter coupling term that arose from the gauge transformation (Eq.~\eqref{eq:PZW_Hph}), one obtains a compact expression for the light--matter interaction \cite{buhmann2013dispersion_book}, 
\begin{equation}
    \hat{V}_\mathrm{int} = - \sum_i\int \dd \mathbf{r} \,\hat{\mathbf{P}}_i (\mathbf{r}) \cdot \hat{\mathbf{E}} (\mathbf{r});
\end{equation}
extracting terms from this and canceling them will only complicate this expression.

For completeness, we close this section by giving the complete Hamiltonian after the PZW transformation and taking the LWA:
\begin{equation} \label{eq:HPZW_LWA}
\begin{aligned}
\hat{H}&_\mathrm{ED}^\mathrm{mQED} =  \hat{H}_\mathrm{ph} + \sum_{i=A,B}\bigg[\sum_{\alpha \in i} \frac{\hat{\mathbf{p}}_\alpha^2}{2m_\alpha} + \hat{V}_{\mathrm{Coulomb},i}   \\
&- \hat{\boldsymbol{\mu}}_i  (\mathbf{r}) \cdot \hat{\mathbf{E}} (\mathbf{r}_i)+\sum_{\lambda}\int\frac{\dd^3\mathbf{k}}{(2\pi)^2}\, (\mathbf{e}_{\mathbf{k},\lambda}\cdot \hat{\boldsymbol{\mu}}_i)^2 \bigg].\\
    \end{aligned}
\end{equation}
To construct $\hat{\mathbf{E}} (\mathbf{r}_i)$, one needs the dyadic Green's function; for our Fabry--P\'erot geometry, this is can for example be found in Ref.~\cite{buhmann2013dispersion_book}.

\section{Discussion and conclusions}
In this work, we have revisited how the intermolecular Coulombic and dipole self-energy interactions cancel in free space, {and discussed under which conditions this holds}. We have also shown how this extends to the cavity-mediated interaction in a perfect Fabry--P\'erot cavity, and provided some numerical evidence suggesting that the cancellation may hold in imperfect cavities as well. Finally, we have shown how this discussion fits into the more general framework of macroscopic QED. 

The most pressing question now is, what are the implications of this for the single-mode Hamiltonian commonly used to study chemical reactivity in microcavities? As it turns out, {if one wants to obtain an expression that does not contain  intermolecular Coulomb interactions (as in Eq.~\eqref{eq:singlemode} for example)}, one can argue two ways, depending on at which stage one takes the single-mode approximation (dropping the sum over $\mathbf{k}$ and $\lambda$, to retain only a single, ``most important" mode). 
Starting in the electric-dipole gauge:
\begin{enumerate}
    \item One can first take the single-mode approximation, and retain a single light--matter coupling term as well as its corresponding DSE term. After this, one can neglect the intermolecular Coulombic interactions: if the molecules are far apart, the direct interactions are small in magnitude, and unless the molecules are very close to the mirrors, the cavity-mediated (image) interactions are even smaller. This then yields the Hamiltonian in the form it is often used (\emph{cf.}~Eq.~\eqref{eq:singlemode}), \emph{i.e.} including DSE cross terms.
    \item One can first cancel the free-space-like intermolecular Coulombic and DSE interactions, and cancel (or neglect the remainder of) the cavity-mediated contributions. Taking the single-mode approximation after this then yields a Hamiltonian nearly identical to Eq.~\eqref{eq:singlemode}, but now \emph{excluding} the DSE cross terms.
\end{enumerate}

It is perhaps not surprising that the order of transformations and approximations matters; this is somewhat analogous to how truncation before or after a gauge transformation leads to different results \cite{taylor2020resolution,stokes2022implications}. 
The two approaches above seem equally valid, and there is no real reason to \emph{a priori} favor one over the other. 
This may seem problematic, however one should keep in mind that one should in any case be careful if the DSE cross terms turn out to play a crucial role in one's theory: as the DSE is a non-resonant term, there is no reason for the single-mode approximation to be justified in that case, and one should likely go beyond it.

In any event, it seems deceptive to label the DSE cross terms as encoding a distance-independent interaction. This apparent distance-independence only arises after taking the single-mode approximation, as one thereby essentially  picks out a single Fourier component. By keeping the sum over $\mathbf{k}$, one obviously recovers the distance-dependence. 

It is important to remark that although the DSE cross terms can be shown to cancel under certain conditions, the DSE self-terms remain. This justifies the inclusion of the DSE term in cavity-coupled electronic structure methods, such as QEDFT or QED-CC \cite{ruggenthaler2023understanding,castagnola2024polaritonic}: in a calculation involving multiple molecules coupled to the same cavity mode, the molecules are often treated as a single ``super-molecule" \cite{schnappinger2023PES,castagnola2024collective,ruggenthaler2023understanding,huang2025local}. In such a treatment, the long-wavelength approximation is usually made with respect to the center of the super-molecule, rather than with respect to the center of each molecule individually (\emph{cf.}~Eq.~\eqref{eq:gaugeU-free}). This results in DSE terms that are purely self-terms, and therefore cannot be canceled out. (Do note that in a sense, this approach implicitly assumes that all molecules are positioned very close together, which is may not necessarily be the case.)
Moreover, a calculation involving multiple molecules coupled to the same cavity mode includes all Coulombic interactions by default; the intermolecular Coulombic interactions are therefore automatically included, meaning that there is no reason to discard part of the DSE term. 

To investigate which of the two single-mode Hamiltonians described above is the most appropriate, 
one could of course try to benchmark results 
against the full Hamiltonian, in which all modes are kept. However, directly performing calculations based on the full Hamiltonian is not just expensive, it may also return diverging results. This is a well-known feature of free-space QED \cite{ruggenthaler2023understanding,scheel2009macroscopic,milonni2013quantum}, and is for example manifested by the divergence of the Lamb shift. Formally, it is remedied by mass renormalization: the mass in the Pauli--Fierz Hamiltonian should be interpreted as the bare mass, and only after accounting for the interaction with the electromagnetic continuum does the physical mass emerge. Therefore, by taking the single-mode approximation, one does not just discard all but one of the terms in the sum over modes; one also effectively subsumes the continuum into the bare mass, so that one can work with the physical mass. 
It should be noted that more rigorous approaches to taking a single/few-mode approximation exist \cite{medina2021few,sanchez2022few,chuang2024microscopic}, where for example the diverging free-space part of the Lamb shift is isolated and dropped \cite{chuang2024microscopic}. However, these formulations seem to have \emph{a priori} excluded the (self-)DSE contributions, which in fact also diverge \cite{cohen1997photons}. It would be interesting to investigate how to rigorously eliminate all of these diverging terms to obtain a Hamiltonian as postulated in Ref.~\cite{svendsen2023theory}, where the free-space part of the transverse vector potential is subtracted, and accounted for by using the physical mass. Such a Hamiltonian would open up the way for a comparison between the aforementioned approaches 1 and 2, and more generally, allow for non-perturbative calculations beyond the single-mode approximation.

\section*{Supplementary Material}
The supplementary material contains:
explicit expressions for the polarization vectors in spherical coordinates; the analytical evaluation of the DSE and Coulombic dipole-dipole integrals in free space; the unitary rotation performed to make the light--matter interaction proportional to $\hat{q}_{\mathbf{k},\lambda}$; details on the mode-function based analysis of the imperfect cavity case, both with an analytic and a numerical approach; and additional numerical results as a function of cavity finesse. 

The code to generate the data for the imperfect cavity case is available on \url{https://github.com/maritfiechter/comparedipolecrossterms}.

\section*{Acknowledgments}
The authors would like to sincerely thank Jeremy Richardson for many helpful discussions and careful proofreading, and Michael Ruggenthaler for insightful discussions. M.R.F. was supported by an
ETH Zurich Research Grant. M.K.S. is supported by the Novo Nordisk Foundation, Grant number NNF22SA0081175, NNF Quantum Computing Programme. 

\section*{Author Declarations}
\subsection*{Conflict of Interest}
The authors have no conflicts to disclose.

\subsection*{Author Contributions}

\textbf{Marit R. Fiechter:} Conceptualization (lead); Formal Analysis (lead); Investigation (lead);  Methodology (equal); Software (lead); Writing -- original draft (lead); Writing -- review and editing (equal).
\textbf{Mark Kamper Svendsen:} Methodology (equal);
Writing -- review and editing (equal).

\section*{Data Availability}
The data that support the findings of this study are available from the corresponding author upon reasonable request.

\bibliography{ref}

\begin{thebibliography}{96}%
\makeatletter
\providecommand \@ifxundefined [1]{%
 \@ifx{#1\undefined}
}%
\providecommand \@ifnum [1]{%
 \ifnum #1\expandafter \@firstoftwo
 \else \expandafter \@secondoftwo
 \fi
}%
\providecommand \@ifx [1]{%
 \ifx #1\expandafter \@firstoftwo
 \else \expandafter \@secondoftwo
 \fi
}%
\providecommand \natexlab [1]{#1}%
\providecommand \enquote  [1]{``#1''}%
\providecommand \bibnamefont  [1]{#1}%
\providecommand \bibfnamefont [1]{#1}%
\providecommand \citenamefont [1]{#1}%
\providecommand \href@noop [0]{\@secondoftwo}%
\providecommand \href [0]{\begingroup \@sanitize@url \@href}%
\providecommand \@href[1]{\@@startlink{#1}\@@href}%
\providecommand \@@href[1]{\endgroup#1\@@endlink}%
\providecommand \@sanitize@url [0]{\catcode `\\12\catcode `\$12\catcode `\&12\catcode `\#12\catcode `\^12\catcode `\_12\catcode `\%12\relax}%
\providecommand \@@startlink[1]{}%
\providecommand \@@endlink[0]{}%
\providecommand \url  [0]{\begingroup\@sanitize@url \@url }%
\providecommand \@url [1]{\endgroup\@href {#1}{\urlprefix }}%
\providecommand \urlprefix  [0]{URL }%
\providecommand \Eprint [0]{\href }%
\providecommand \doibase [0]{https://doi.org/}%
\providecommand \selectlanguage [0]{\@gobble}%
\providecommand \bibinfo  [0]{\@secondoftwo}%
\providecommand \bibfield  [0]{\@secondoftwo}%
\providecommand \translation [1]{[#1]}%
\providecommand \BibitemOpen [0]{}%
\providecommand \bibitemStop [0]{}%
\providecommand \bibitemNoStop [0]{.\EOS\space}%
\providecommand \EOS [0]{\spacefactor3000\relax}%
\providecommand \BibitemShut  [1]{\csname bibitem#1\endcsname}%
\let\auto@bib@innerbib\@empty
\bibitem [{\citenamefont {Nagarajan}\ \emph {et~al.}(2021)\citenamefont {Nagarajan}, \citenamefont {Thomas},\ and\ \citenamefont {Ebbesen}}]{nagarajan2021chemistry}%
  \BibitemOpen
  \bibfield  {author} {\bibinfo {author} {\bibfnamefont {K.}~\bibnamefont {Nagarajan}}, \bibinfo {author} {\bibfnamefont {A.}~\bibnamefont {Thomas}},\ and\ \bibinfo {author} {\bibfnamefont {T.~W.}\ \bibnamefont {Ebbesen}},\ }\bibfield  {title} {\bibinfo {title} {Chemistry under vibrational strong coupling},\ }\href@noop {} {\bibfield  {journal} {\bibinfo  {journal} {J.~Am. Chem. Soc.}\ }\textbf {\bibinfo {volume} {143}},\ \bibinfo {pages} {16877} (\bibinfo {year} {2021})}\BibitemShut {NoStop}%
\bibitem [{\citenamefont {Dunkelberger}\ \emph {et~al.}(2022)\citenamefont {Dunkelberger}, \citenamefont {Simpkins}, \citenamefont {Vurgaftman},\ and\ \citenamefont {Owrutsky}}]{dunkelberger2022vibration}%
  \BibitemOpen
  \bibfield  {author} {\bibinfo {author} {\bibfnamefont {A.~D.}\ \bibnamefont {Dunkelberger}}, \bibinfo {author} {\bibfnamefont {B.~S.}\ \bibnamefont {Simpkins}}, \bibinfo {author} {\bibfnamefont {I.}~\bibnamefont {Vurgaftman}},\ and\ \bibinfo {author} {\bibfnamefont {J.~C.}\ \bibnamefont {Owrutsky}},\ }\bibfield  {title} {\bibinfo {title} {Vibration-cavity polariton chemistry and dynamics},\ }\href@noop {} {\bibfield  {journal} {\bibinfo  {journal} {Annu. Rev. Phys. Chem.}\ }\textbf {\bibinfo {volume} {73}},\ \bibinfo {pages} {429} (\bibinfo {year} {2022})}\BibitemShut {NoStop}%
\bibitem [{\citenamefont {Thomas}\ \emph {et~al.}(2016)\citenamefont {Thomas}, \citenamefont {George}, \citenamefont {Shalabney}, \citenamefont {Dryzhakov}, \citenamefont {Varma}, \citenamefont {Moran}, \citenamefont {Chervy}, \citenamefont {Zhong}, \citenamefont {Devaux}, \citenamefont {Genet}, \citenamefont {Hutchison},\ and\ \citenamefont {Ebbesen}}]{thomas2016ground}%
  \BibitemOpen
  \bibfield  {author} {\bibinfo {author} {\bibfnamefont {A.}~\bibnamefont {Thomas}}, \bibinfo {author} {\bibfnamefont {J.}~\bibnamefont {George}}, \bibinfo {author} {\bibfnamefont {A.}~\bibnamefont {Shalabney}}, \bibinfo {author} {\bibfnamefont {M.}~\bibnamefont {Dryzhakov}}, \bibinfo {author} {\bibfnamefont {S.~J.}\ \bibnamefont {Varma}}, \bibinfo {author} {\bibfnamefont {J.}~\bibnamefont {Moran}}, \bibinfo {author} {\bibfnamefont {T.}~\bibnamefont {Chervy}}, \bibinfo {author} {\bibfnamefont {X.}~\bibnamefont {Zhong}}, \bibinfo {author} {\bibfnamefont {E.}~\bibnamefont {Devaux}}, \bibinfo {author} {\bibfnamefont {C.}~\bibnamefont {Genet}}, \bibinfo {author} {\bibfnamefont {J.~A.}\ \bibnamefont {Hutchison}},\ and\ \bibinfo {author} {\bibfnamefont {T.~W.}\ \bibnamefont {Ebbesen}},\ }\bibfield  {title} {\bibinfo {title} {Ground-state chemical reactivity under vibrational coupling to the vacuum electromagnetic field},\ }\href@noop {} {\bibfield  {journal} {\bibinfo  {journal} {Angew. Chem. Int. Edit.}\ }\textbf
  {\bibinfo {volume} {55}},\ \bibinfo {pages} {11462} (\bibinfo {year} {2016})}\BibitemShut {NoStop}%
\bibitem [{\citenamefont {Thomas}\ \emph {et~al.}(2019)\citenamefont {Thomas}, \citenamefont {Lethuillier-Karl}, \citenamefont {Nagarajan}, \citenamefont {Vergauwe}, \citenamefont {George}, \citenamefont {Chervy}, \citenamefont {Shalabney}, \citenamefont {Devaux}, \citenamefont {Genet}, \citenamefont {Moran},\ and\ \citenamefont {Ebbesen}}]{thomas2019tilting}%
  \BibitemOpen
  \bibfield  {author} {\bibinfo {author} {\bibfnamefont {A.}~\bibnamefont {Thomas}}, \bibinfo {author} {\bibfnamefont {L.}~\bibnamefont {Lethuillier-Karl}}, \bibinfo {author} {\bibfnamefont {K.}~\bibnamefont {Nagarajan}}, \bibinfo {author} {\bibfnamefont {R.~M.~A.}\ \bibnamefont {Vergauwe}}, \bibinfo {author} {\bibfnamefont {J.}~\bibnamefont {George}}, \bibinfo {author} {\bibfnamefont {T.}~\bibnamefont {Chervy}}, \bibinfo {author} {\bibfnamefont {A.}~\bibnamefont {Shalabney}}, \bibinfo {author} {\bibfnamefont {E.}~\bibnamefont {Devaux}}, \bibinfo {author} {\bibfnamefont {C.}~\bibnamefont {Genet}}, \bibinfo {author} {\bibfnamefont {J.}~\bibnamefont {Moran}},\ and\ \bibinfo {author} {\bibfnamefont {T.~W.}\ \bibnamefont {Ebbesen}},\ }\bibfield  {title} {\bibinfo {title} {Tilting a ground-state reactivity landscape by vibrational strong coupling},\ }\href@noop {} {\bibfield  {journal} {\bibinfo  {journal} {Science}\ }\textbf {\bibinfo {volume} {363}},\ \bibinfo {pages} {615} (\bibinfo {year} {2019})}\BibitemShut
  {NoStop}%
\bibitem [{\citenamefont {Lather}\ \emph {et~al.}(2019)\citenamefont {Lather}, \citenamefont {Bhatt}, \citenamefont {Thomas}, \citenamefont {Ebbesen},\ and\ \citenamefont {George}}]{lather2019cavity}%
  \BibitemOpen
  \bibfield  {author} {\bibinfo {author} {\bibfnamefont {J.}~\bibnamefont {Lather}}, \bibinfo {author} {\bibfnamefont {P.}~\bibnamefont {Bhatt}}, \bibinfo {author} {\bibfnamefont {A.}~\bibnamefont {Thomas}}, \bibinfo {author} {\bibfnamefont {T.~W.}\ \bibnamefont {Ebbesen}},\ and\ \bibinfo {author} {\bibfnamefont {J.}~\bibnamefont {George}},\ }\bibfield  {title} {\bibinfo {title} {Cavity catalysis by cooperative vibrational strong coupling of reactant and solvent molecules},\ }\href@noop {} {\bibfield  {journal} {\bibinfo  {journal} {Angew. Chem. Int. Edit.}\ }\textbf {\bibinfo {volume} {58}},\ \bibinfo {pages} {10635} (\bibinfo {year} {2019})}\BibitemShut {NoStop}%
\bibitem [{\citenamefont {Vergauwe}\ \emph {et~al.}(2019)\citenamefont {Vergauwe}, \citenamefont {Thomas}, \citenamefont {Nagarajan}, \citenamefont {Shalabney}, \citenamefont {George}, \citenamefont {Chervy}, \citenamefont {Seidel}, \citenamefont {Devaux}, \citenamefont {Torbeev},\ and\ \citenamefont {Ebbesen}}]{vergauwe2019modification}%
  \BibitemOpen
  \bibfield  {author} {\bibinfo {author} {\bibfnamefont {R.~M.}\ \bibnamefont {Vergauwe}}, \bibinfo {author} {\bibfnamefont {A.}~\bibnamefont {Thomas}}, \bibinfo {author} {\bibfnamefont {K.}~\bibnamefont {Nagarajan}}, \bibinfo {author} {\bibfnamefont {A.}~\bibnamefont {Shalabney}}, \bibinfo {author} {\bibfnamefont {J.}~\bibnamefont {George}}, \bibinfo {author} {\bibfnamefont {T.}~\bibnamefont {Chervy}}, \bibinfo {author} {\bibfnamefont {M.}~\bibnamefont {Seidel}}, \bibinfo {author} {\bibfnamefont {E.}~\bibnamefont {Devaux}}, \bibinfo {author} {\bibfnamefont {V.}~\bibnamefont {Torbeev}},\ and\ \bibinfo {author} {\bibfnamefont {T.~W.}\ \bibnamefont {Ebbesen}},\ }\bibfield  {title} {\bibinfo {title} {Modification of enzyme activity by vibrational strong coupling of water},\ }\href@noop {} {\bibfield  {journal} {\bibinfo  {journal} {Angew. Chem. Int. Edit.}\ }\textbf {\bibinfo {volume} {58}},\ \bibinfo {pages} {15324} (\bibinfo {year} {2019})}\BibitemShut {NoStop}%
\bibitem [{\citenamefont {Hirai}\ \emph {et~al.}(2020)\citenamefont {Hirai}, \citenamefont {Takeda}, \citenamefont {Hutchison},\ and\ \citenamefont {Uji-i}}]{hirai2020modulation}%
  \BibitemOpen
  \bibfield  {author} {\bibinfo {author} {\bibfnamefont {K.}~\bibnamefont {Hirai}}, \bibinfo {author} {\bibfnamefont {R.}~\bibnamefont {Takeda}}, \bibinfo {author} {\bibfnamefont {J.~A.}\ \bibnamefont {Hutchison}},\ and\ \bibinfo {author} {\bibfnamefont {H.}~\bibnamefont {Uji-i}},\ }\bibfield  {title} {\bibinfo {title} {Modulation of {P}rins cyclization by vibrational strong coupling},\ }\href@noop {} {\bibfield  {journal} {\bibinfo  {journal} {Angew. Chem. Int. Edit.}\ }\textbf {\bibinfo {volume} {59}},\ \bibinfo {pages} {5332} (\bibinfo {year} {2020})}\BibitemShut {NoStop}%
\bibitem [{\citenamefont {Sau}\ \emph {et~al.}(2021)\citenamefont {Sau}, \citenamefont {Nagarajan}, \citenamefont {Patrahau}, \citenamefont {Lethuillier-Karl}, \citenamefont {Vergauwe}, \citenamefont {Thomas}, \citenamefont {Moran}, \citenamefont {Genet},\ and\ \citenamefont {Ebbesen}}]{sau2021modifying}%
  \BibitemOpen
  \bibfield  {author} {\bibinfo {author} {\bibfnamefont {A.}~\bibnamefont {Sau}}, \bibinfo {author} {\bibfnamefont {K.}~\bibnamefont {Nagarajan}}, \bibinfo {author} {\bibfnamefont {B.}~\bibnamefont {Patrahau}}, \bibinfo {author} {\bibfnamefont {L.}~\bibnamefont {Lethuillier-Karl}}, \bibinfo {author} {\bibfnamefont {R.~M.}\ \bibnamefont {Vergauwe}}, \bibinfo {author} {\bibfnamefont {A.}~\bibnamefont {Thomas}}, \bibinfo {author} {\bibfnamefont {J.}~\bibnamefont {Moran}}, \bibinfo {author} {\bibfnamefont {C.}~\bibnamefont {Genet}},\ and\ \bibinfo {author} {\bibfnamefont {T.~W.}\ \bibnamefont {Ebbesen}},\ }\bibfield  {title} {\bibinfo {title} {Modifying {Woodward--Hoffmann} stereoselectivity under vibrational strong coupling},\ }\href@noop {} {\bibfield  {journal} {\bibinfo  {journal} {Angew. Chem. Int. Edit.}\ }\textbf {\bibinfo {volume} {60}},\ \bibinfo {pages} {5712} (\bibinfo {year} {2021})}\BibitemShut {NoStop}%
\bibitem [{\citenamefont {Ahn}\ \emph {et~al.}(2023)\citenamefont {Ahn}, \citenamefont {Triana}, \citenamefont {Recabal}, \citenamefont {Herrera},\ and\ \citenamefont {Simpkins}}]{ahn2023modification}%
  \BibitemOpen
  \bibfield  {author} {\bibinfo {author} {\bibfnamefont {W.}~\bibnamefont {Ahn}}, \bibinfo {author} {\bibfnamefont {J.~F.}\ \bibnamefont {Triana}}, \bibinfo {author} {\bibfnamefont {F.}~\bibnamefont {Recabal}}, \bibinfo {author} {\bibfnamefont {F.}~\bibnamefont {Herrera}},\ and\ \bibinfo {author} {\bibfnamefont {B.~S.}\ \bibnamefont {Simpkins}},\ }\bibfield  {title} {\bibinfo {title} {Modification of ground-state chemical reactivity via light--matter coherence in infrared cavities},\ }\href@noop {} {\bibfield  {journal} {\bibinfo  {journal} {Science}\ }\textbf {\bibinfo {volume} {380}},\ \bibinfo {pages} {1165} (\bibinfo {year} {2023})}\BibitemShut {NoStop}%
\bibitem [{\citenamefont {Pang}\ \emph {et~al.}(2020)\citenamefont {Pang}, \citenamefont {Thomas}, \citenamefont {Nagarajan}, \citenamefont {Vergauwe}, \citenamefont {Joseph}, \citenamefont {Patrahau}, \citenamefont {Wang}, \citenamefont {Genet},\ and\ \citenamefont {Ebbesen}}]{pang2020role}%
  \BibitemOpen
  \bibfield  {author} {\bibinfo {author} {\bibfnamefont {Y.}~\bibnamefont {Pang}}, \bibinfo {author} {\bibfnamefont {A.}~\bibnamefont {Thomas}}, \bibinfo {author} {\bibfnamefont {K.}~\bibnamefont {Nagarajan}}, \bibinfo {author} {\bibfnamefont {R.~M.}\ \bibnamefont {Vergauwe}}, \bibinfo {author} {\bibfnamefont {K.}~\bibnamefont {Joseph}}, \bibinfo {author} {\bibfnamefont {B.}~\bibnamefont {Patrahau}}, \bibinfo {author} {\bibfnamefont {K.}~\bibnamefont {Wang}}, \bibinfo {author} {\bibfnamefont {C.}~\bibnamefont {Genet}},\ and\ \bibinfo {author} {\bibfnamefont {T.~W.}\ \bibnamefont {Ebbesen}},\ }\bibfield  {title} {\bibinfo {title} {On the role of symmetry in vibrational strong coupling: the case of charge-transfer complexation},\ }\href@noop {} {\bibfield  {journal} {\bibinfo  {journal} {Angew. Chem. Int. Edit.}\ }\textbf {\bibinfo {volume} {59}},\ \bibinfo {pages} {10436} (\bibinfo {year} {2020})}\BibitemShut {NoStop}%
\bibitem [{\citenamefont {Patrahau}\ \emph {et~al.}(2024)\citenamefont {Patrahau}, \citenamefont {Piejko}, \citenamefont {Mayer}, \citenamefont {Antheaume}, \citenamefont {Sangchai}, \citenamefont {Ragazzon}, \citenamefont {Jayachandran}, \citenamefont {Devaux}, \citenamefont {Genet}, \citenamefont {Moran} \emph {et~al.}}]{patrahau2024direct}%
  \BibitemOpen
  \bibfield  {author} {\bibinfo {author} {\bibfnamefont {B.}~\bibnamefont {Patrahau}}, \bibinfo {author} {\bibfnamefont {M.}~\bibnamefont {Piejko}}, \bibinfo {author} {\bibfnamefont {R.~J.}\ \bibnamefont {Mayer}}, \bibinfo {author} {\bibfnamefont {C.}~\bibnamefont {Antheaume}}, \bibinfo {author} {\bibfnamefont {T.}~\bibnamefont {Sangchai}}, \bibinfo {author} {\bibfnamefont {G.}~\bibnamefont {Ragazzon}}, \bibinfo {author} {\bibfnamefont {A.}~\bibnamefont {Jayachandran}}, \bibinfo {author} {\bibfnamefont {E.}~\bibnamefont {Devaux}}, \bibinfo {author} {\bibfnamefont {C.}~\bibnamefont {Genet}}, \bibinfo {author} {\bibfnamefont {J.}~\bibnamefont {Moran}}, \emph {et~al.},\ }\bibfield  {title} {\bibinfo {title} {Direct observation of polaritonic chemistry by nuclear magnetic resonance spectroscopy},\ }\href@noop {} {\bibfield  {journal} {\bibinfo  {journal} {Angew. Chem. Int. Edit.}\ }\textbf {\bibinfo {volume} {63}},\ \bibinfo {pages} {e202401368} (\bibinfo {year} {2024})}\BibitemShut {NoStop}%
\bibitem [{\citenamefont {Sandeep}\ \emph {et~al.}(2022)\citenamefont {Sandeep}, \citenamefont {Joseph}, \citenamefont {Gautier}, \citenamefont {Nagarajan}, \citenamefont {Sujith}, \citenamefont {Thomas},\ and\ \citenamefont {Ebbesen}}]{sandeep2022manipulating}%
  \BibitemOpen
  \bibfield  {author} {\bibinfo {author} {\bibfnamefont {K.}~\bibnamefont {Sandeep}}, \bibinfo {author} {\bibfnamefont {K.}~\bibnamefont {Joseph}}, \bibinfo {author} {\bibfnamefont {J.}~\bibnamefont {Gautier}}, \bibinfo {author} {\bibfnamefont {K.}~\bibnamefont {Nagarajan}}, \bibinfo {author} {\bibfnamefont {M.}~\bibnamefont {Sujith}}, \bibinfo {author} {\bibfnamefont {K.~G.}\ \bibnamefont {Thomas}},\ and\ \bibinfo {author} {\bibfnamefont {T.~W.}\ \bibnamefont {Ebbesen}},\ }\bibfield  {title} {\bibinfo {title} {Manipulating the self-assembly of phenyleneethynylenes under vibrational strong coupling},\ }\href@noop {} {\bibfield  {journal} {\bibinfo  {journal} {J. Phys. Chem. Lett.}\ }\textbf {\bibinfo {volume} {13}},\ \bibinfo {pages} {1209} (\bibinfo {year} {2022})}\BibitemShut {NoStop}%
\bibitem [{\citenamefont {Imai}\ \emph {et~al.}(2025)\citenamefont {Imai}, \citenamefont {Hamada}, \citenamefont {Nozaki}, \citenamefont {Fujita}, \citenamefont {Takahashi}, \citenamefont {Fujita}, \citenamefont {Harano}, \citenamefont {Uji-i}, \citenamefont {Takai},\ and\ \citenamefont {Hirai}}]{imai2025accessing}%
  \BibitemOpen
  \bibfield  {author} {\bibinfo {author} {\bibfnamefont {S.}~\bibnamefont {Imai}}, \bibinfo {author} {\bibfnamefont {T.}~\bibnamefont {Hamada}}, \bibinfo {author} {\bibfnamefont {M.}~\bibnamefont {Nozaki}}, \bibinfo {author} {\bibfnamefont {T.}~\bibnamefont {Fujita}}, \bibinfo {author} {\bibfnamefont {M.}~\bibnamefont {Takahashi}}, \bibinfo {author} {\bibfnamefont {Y.}~\bibnamefont {Fujita}}, \bibinfo {author} {\bibfnamefont {K.}~\bibnamefont {Harano}}, \bibinfo {author} {\bibfnamefont {H.}~\bibnamefont {Uji-i}}, \bibinfo {author} {\bibfnamefont {A.}~\bibnamefont {Takai}},\ and\ \bibinfo {author} {\bibfnamefont {K.}~\bibnamefont {Hirai}},\ }\bibfield  {title} {\bibinfo {title} {Accessing a hidden pathway to supramolecular toroid through vibrational strong coupling},\ }\href@noop {} {\bibfield  {journal} {\bibinfo  {journal} {J. Am. Chem. Soc.}\ } (\bibinfo {year} {2025})}\BibitemShut {NoStop}%
\bibitem [{\citenamefont {George}\ and\ \citenamefont {Singh}(2023)}]{george2023polaritonic}%
  \BibitemOpen
  \bibfield  {author} {\bibinfo {author} {\bibfnamefont {J.}~\bibnamefont {George}}\ and\ \bibinfo {author} {\bibfnamefont {J.}~\bibnamefont {Singh}},\ }\bibfield  {title} {\bibinfo {title} {Polaritonic chemistry: Band-selective control of chemical reactions by vibrational strong coupling},\ }\href@noop {} {\bibfield  {journal} {\bibinfo  {journal} {ACS Catal.}\ }\textbf {\bibinfo {volume} {13}},\ \bibinfo {pages} {2631} (\bibinfo {year} {2023})}\BibitemShut {NoStop}%
\bibitem [{\citenamefont {Campos-Gonzalez-Angulo}\ \emph {et~al.}(2023)\citenamefont {Campos-Gonzalez-Angulo}, \citenamefont {Poh}, \citenamefont {Du},\ and\ \citenamefont {Yuen-Zhou}}]{campos2023swinging}%
  \BibitemOpen
  \bibfield  {author} {\bibinfo {author} {\bibfnamefont {J.~A.}\ \bibnamefont {Campos-Gonzalez-Angulo}}, \bibinfo {author} {\bibfnamefont {Y.~R.}\ \bibnamefont {Poh}}, \bibinfo {author} {\bibfnamefont {M.}~\bibnamefont {Du}},\ and\ \bibinfo {author} {\bibfnamefont {J.}~\bibnamefont {Yuen-Zhou}},\ }\bibfield  {title} {\bibinfo {title} {{Swinging between shine and shadow: Theoretical advances on thermally activated vibropolaritonic chemistry}},\ }\href@noop {} {\bibfield  {journal} {\bibinfo  {journal} {J. Chem. Phys.}\ }\textbf {\bibinfo {volume} {158}},\ \bibinfo {pages} {230901} (\bibinfo {year} {2023})}\BibitemShut {NoStop}%
\bibitem [{\citenamefont {Mandal}\ \emph {et~al.}(2023{\natexlab{a}})\citenamefont {Mandal}, \citenamefont {Taylor}, \citenamefont {Weight}, \citenamefont {Koessler}, \citenamefont {Li},\ and\ \citenamefont {Huo}}]{mandal2023theoretical}%
  \BibitemOpen
  \bibfield  {author} {\bibinfo {author} {\bibfnamefont {A.}~\bibnamefont {Mandal}}, \bibinfo {author} {\bibfnamefont {M.~A.}\ \bibnamefont {Taylor}}, \bibinfo {author} {\bibfnamefont {B.~M.}\ \bibnamefont {Weight}}, \bibinfo {author} {\bibfnamefont {E.~R.}\ \bibnamefont {Koessler}}, \bibinfo {author} {\bibfnamefont {X.}~\bibnamefont {Li}},\ and\ \bibinfo {author} {\bibfnamefont {P.}~\bibnamefont {Huo}},\ }\bibfield  {title} {\bibinfo {title} {Theoretical advances in polariton chemistry and molecular cavity quantum electrodynamics},\ }\href@noop {} {\bibfield  {journal} {\bibinfo  {journal} {Chem. Rev.}\ }\textbf {\bibinfo {volume} {123}},\ \bibinfo {pages} {9786} (\bibinfo {year} {2023}{\natexlab{a}})}\BibitemShut {NoStop}%
\bibitem [{\citenamefont {Ruggenthaler}\ \emph {et~al.}(2023)\citenamefont {Ruggenthaler}, \citenamefont {Sidler},\ and\ \citenamefont {Rubio}}]{ruggenthaler2023understanding}%
  \BibitemOpen
  \bibfield  {author} {\bibinfo {author} {\bibfnamefont {M.}~\bibnamefont {Ruggenthaler}}, \bibinfo {author} {\bibfnamefont {D.}~\bibnamefont {Sidler}},\ and\ \bibinfo {author} {\bibfnamefont {A.}~\bibnamefont {Rubio}},\ }\bibfield  {title} {\bibinfo {title} {Understanding polaritonic chemistry from ab initio quantum electrodynamics},\ }\href@noop {} {\bibfield  {journal} {\bibinfo  {journal} {Chem. Rev.}\ }\textbf {\bibinfo {volume} {123}},\ \bibinfo {pages} {11191} (\bibinfo {year} {2023})}\BibitemShut {NoStop}%
\bibitem [{\citenamefont {Galego}\ \emph {et~al.}(2019)\citenamefont {Galego}, \citenamefont {Climent}, \citenamefont {Garcia-Vidal},\ and\ \citenamefont {Feist}}]{galego2019cavity}%
  \BibitemOpen
  \bibfield  {author} {\bibinfo {author} {\bibfnamefont {J.}~\bibnamefont {Galego}}, \bibinfo {author} {\bibfnamefont {C.}~\bibnamefont {Climent}}, \bibinfo {author} {\bibfnamefont {F.~J.}\ \bibnamefont {Garcia-Vidal}},\ and\ \bibinfo {author} {\bibfnamefont {J.}~\bibnamefont {Feist}},\ }\bibfield  {title} {\bibinfo {title} {Cavity {Casimir-Polder} forces and their effects in ground-state chemical reactivity},\ }\href@noop {} {\bibfield  {journal} {\bibinfo  {journal} {Phys. Rev. X}\ }\textbf {\bibinfo {volume} {9}},\ \bibinfo {pages} {021057} (\bibinfo {year} {2019})}\BibitemShut {NoStop}%
\bibitem [{\citenamefont {Li}\ \emph {et~al.}(2020)\citenamefont {Li}, \citenamefont {Nitzan},\ and\ \citenamefont {Subotnik}}]{li2020origin}%
  \BibitemOpen
  \bibfield  {author} {\bibinfo {author} {\bibfnamefont {T.~E.}\ \bibnamefont {Li}}, \bibinfo {author} {\bibfnamefont {A.}~\bibnamefont {Nitzan}},\ and\ \bibinfo {author} {\bibfnamefont {J.~E.}\ \bibnamefont {Subotnik}},\ }\bibfield  {title} {\bibinfo {title} {On the origin of ground-state vacuum-field catalysis: Equilibrium consideration},\ }\href@noop {} {\bibfield  {journal} {\bibinfo  {journal} {J. Chem. Phys.}\ }\textbf {\bibinfo {volume} {152}} (\bibinfo {year} {2020})}\BibitemShut {NoStop}%
\bibitem [{\citenamefont {Li}\ \emph {et~al.}(2021)\citenamefont {Li}, \citenamefont {Mandal},\ and\ \citenamefont {Huo}}]{li2021cavity}%
  \BibitemOpen
  \bibfield  {author} {\bibinfo {author} {\bibfnamefont {X.}~\bibnamefont {Li}}, \bibinfo {author} {\bibfnamefont {A.}~\bibnamefont {Mandal}},\ and\ \bibinfo {author} {\bibfnamefont {P.}~\bibnamefont {Huo}},\ }\bibfield  {title} {\bibinfo {title} {Cavity frequency-dependent theory for vibrational polariton chemistry},\ }\href@noop {} {\bibfield  {journal} {\bibinfo  {journal} {Nat. Commun.}\ }\textbf {\bibinfo {volume} {12}},\ \bibinfo {pages} {1315} (\bibinfo {year} {2021})}\BibitemShut {NoStop}%
\bibitem [{\citenamefont {Yang}\ and\ \citenamefont {Cao}(2021)}]{yang2021quantum}%
  \BibitemOpen
  \bibfield  {author} {\bibinfo {author} {\bibfnamefont {P.-Y.}\ \bibnamefont {Yang}}\ and\ \bibinfo {author} {\bibfnamefont {J.}~\bibnamefont {Cao}},\ }\bibfield  {title} {\bibinfo {title} {Quantum effects in chemical reactions under polaritonic vibrational strong coupling},\ }\href@noop {} {\bibfield  {journal} {\bibinfo  {journal} {J. Phys. Chem. Lett.}\ }\textbf {\bibinfo {volume} {12}},\ \bibinfo {pages} {9531} (\bibinfo {year} {2021})}\BibitemShut {NoStop}%
\bibitem [{\citenamefont {Lindoy}\ \emph {et~al.}(2022)\citenamefont {Lindoy}, \citenamefont {Mandal},\ and\ \citenamefont {Reichman}}]{lindoy2022resonant}%
  \BibitemOpen
  \bibfield  {author} {\bibinfo {author} {\bibfnamefont {L.~P.}\ \bibnamefont {Lindoy}}, \bibinfo {author} {\bibfnamefont {A.}~\bibnamefont {Mandal}},\ and\ \bibinfo {author} {\bibfnamefont {D.~R.}\ \bibnamefont {Reichman}},\ }\bibfield  {title} {\bibinfo {title} {Resonant cavity modification of ground-state chemical kinetics},\ }\href@noop {} {\bibfield  {journal} {\bibinfo  {journal} {J. Phys. Chem. Lett.}\ }\textbf {\bibinfo {volume} {13}},\ \bibinfo {pages} {6580} (\bibinfo {year} {2022})}\BibitemShut {NoStop}%
\bibitem [{\citenamefont {Wang}\ \emph {et~al.}(2022{\natexlab{a}})\citenamefont {Wang}, \citenamefont {Neuman}, \citenamefont {Yelin},\ and\ \citenamefont {Flick}}]{wang2022cavity}%
  \BibitemOpen
  \bibfield  {author} {\bibinfo {author} {\bibfnamefont {D.~S.}\ \bibnamefont {Wang}}, \bibinfo {author} {\bibfnamefont {T.}~\bibnamefont {Neuman}}, \bibinfo {author} {\bibfnamefont {S.~F.}\ \bibnamefont {Yelin}},\ and\ \bibinfo {author} {\bibfnamefont {J.}~\bibnamefont {Flick}},\ }\bibfield  {title} {\bibinfo {title} {Cavity-modified unimolecular dissociation reactions via intramolecular vibrational energy redistribution},\ }\href@noop {} {\bibfield  {journal} {\bibinfo  {journal} {J. Phys. Chem. Lett.}\ }\textbf {\bibinfo {volume} {13}},\ \bibinfo {pages} {3317} (\bibinfo {year} {2022}{\natexlab{a}})}\BibitemShut {NoStop}%
\bibitem [{\citenamefont {Wang}\ \emph {et~al.}(2022{\natexlab{b}})\citenamefont {Wang}, \citenamefont {Flick},\ and\ \citenamefont {Yelin}}]{wang2022chemical}%
  \BibitemOpen
  \bibfield  {author} {\bibinfo {author} {\bibfnamefont {D.~S.}\ \bibnamefont {Wang}}, \bibinfo {author} {\bibfnamefont {J.}~\bibnamefont {Flick}},\ and\ \bibinfo {author} {\bibfnamefont {S.~F.}\ \bibnamefont {Yelin}},\ }\bibfield  {title} {\bibinfo {title} {Chemical reactivity under collective vibrational strong coupling},\ }\href@noop {} {\bibfield  {journal} {\bibinfo  {journal} {J. Chem. Phys.}\ }\textbf {\bibinfo {volume} {157}} (\bibinfo {year} {2022}{\natexlab{b}})}\BibitemShut {NoStop}%
\bibitem [{\citenamefont {Philbin}\ \emph {et~al.}(2022)\citenamefont {Philbin}, \citenamefont {Wang}, \citenamefont {Narang},\ and\ \citenamefont {Dou}}]{philbin2022chemical}%
  \BibitemOpen
  \bibfield  {author} {\bibinfo {author} {\bibfnamefont {J.~P.}\ \bibnamefont {Philbin}}, \bibinfo {author} {\bibfnamefont {Y.}~\bibnamefont {Wang}}, \bibinfo {author} {\bibfnamefont {P.}~\bibnamefont {Narang}},\ and\ \bibinfo {author} {\bibfnamefont {W.}~\bibnamefont {Dou}},\ }\bibfield  {title} {\bibinfo {title} {Chemical reactions in imperfect cavities: Enhancement, suppression, and resonance},\ }\href@noop {} {\bibfield  {journal} {\bibinfo  {journal} {J. Phys. Chem. C}\ }\textbf {\bibinfo {volume} {126}},\ \bibinfo {pages} {14908} (\bibinfo {year} {2022})}\BibitemShut {NoStop}%
\bibitem [{\citenamefont {Sun}\ and\ \citenamefont {Vendrell}(2022)}]{sun2022suppression}%
  \BibitemOpen
  \bibfield  {author} {\bibinfo {author} {\bibfnamefont {J.}~\bibnamefont {Sun}}\ and\ \bibinfo {author} {\bibfnamefont {O.}~\bibnamefont {Vendrell}},\ }\bibfield  {title} {\bibinfo {title} {Suppression and enhancement of thermal chemical rates in a cavity},\ }\href@noop {} {\bibfield  {journal} {\bibinfo  {journal} {J. Phys. Chem. Lett.}\ }\textbf {\bibinfo {volume} {13}},\ \bibinfo {pages} {4441} (\bibinfo {year} {2022})}\BibitemShut {NoStop}%
\bibitem [{\citenamefont {Sch{\"a}fer}\ \emph {et~al.}(2022)\citenamefont {Sch{\"a}fer}, \citenamefont {Flick}, \citenamefont {Ronca}, \citenamefont {Narang},\ and\ \citenamefont {Rubio}}]{schafer2022shining}%
  \BibitemOpen
  \bibfield  {author} {\bibinfo {author} {\bibfnamefont {C.}~\bibnamefont {Sch{\"a}fer}}, \bibinfo {author} {\bibfnamefont {J.}~\bibnamefont {Flick}}, \bibinfo {author} {\bibfnamefont {E.}~\bibnamefont {Ronca}}, \bibinfo {author} {\bibfnamefont {P.}~\bibnamefont {Narang}},\ and\ \bibinfo {author} {\bibfnamefont {A.}~\bibnamefont {Rubio}},\ }\bibfield  {title} {\bibinfo {title} {Shining light on the microscopic resonant mechanism responsible for cavity-mediated chemical reactivity},\ }\href@noop {} {\bibfield  {journal} {\bibinfo  {journal} {Nat. Commun.}\ }\textbf {\bibinfo {volume} {13}},\ \bibinfo {pages} {7817} (\bibinfo {year} {2022})}\BibitemShut {NoStop}%
\bibitem [{\citenamefont {Du}\ \emph {et~al.}(2023)\citenamefont {Du}, \citenamefont {Poh},\ and\ \citenamefont {Yuen-Zhou}}]{du2023vibropolaritonic}%
  \BibitemOpen
  \bibfield  {author} {\bibinfo {author} {\bibfnamefont {M.}~\bibnamefont {Du}}, \bibinfo {author} {\bibfnamefont {Y.~R.}\ \bibnamefont {Poh}},\ and\ \bibinfo {author} {\bibfnamefont {J.}~\bibnamefont {Yuen-Zhou}},\ }\bibfield  {title} {\bibinfo {title} {Vibropolaritonic reaction rates in the collective strong coupling regime: {Pollak--Grabert--H\"{a}nggi} theory},\ }\href@noop {} {\bibfield  {journal} {\bibinfo  {journal} {J. Phys. Chem. C}\ }\textbf {\bibinfo {volume} {127}},\ \bibinfo {pages} {5230} (\bibinfo {year} {2023})}\BibitemShut {NoStop}%
\bibitem [{\citenamefont {Lindoy}\ \emph {et~al.}(2023)\citenamefont {Lindoy}, \citenamefont {Mandal},\ and\ \citenamefont {Reichman}}]{lindoy2023quantum}%
  \BibitemOpen
  \bibfield  {author} {\bibinfo {author} {\bibfnamefont {L.~P.}\ \bibnamefont {Lindoy}}, \bibinfo {author} {\bibfnamefont {A.}~\bibnamefont {Mandal}},\ and\ \bibinfo {author} {\bibfnamefont {D.~R.}\ \bibnamefont {Reichman}},\ }\bibfield  {title} {\bibinfo {title} {Quantum dynamical effects of vibrational strong coupling in chemical reactivity},\ }\href@noop {} {\bibfield  {journal} {\bibinfo  {journal} {Nat. Commun.}\ }\textbf {\bibinfo {volume} {14}},\ \bibinfo {pages} {2733} (\bibinfo {year} {2023})}\BibitemShut {NoStop}%
\bibitem [{\citenamefont {Fiechter}\ \emph {et~al.}(2023)\citenamefont {Fiechter}, \citenamefont {Runeson}, \citenamefont {Lawrence},\ and\ \citenamefont {Richardson}}]{fiechter2023RPMD}%
  \BibitemOpen
  \bibfield  {author} {\bibinfo {author} {\bibfnamefont {M.~R.}\ \bibnamefont {Fiechter}}, \bibinfo {author} {\bibfnamefont {J.~E.}\ \bibnamefont {Runeson}}, \bibinfo {author} {\bibfnamefont {J.~E.}\ \bibnamefont {Lawrence}},\ and\ \bibinfo {author} {\bibfnamefont {J.~O.}\ \bibnamefont {Richardson}},\ }\bibfield  {title} {\bibinfo {title} {How quantum is the resonance behavior in vibrational polariton chemistry?},\ }\href@noop {} {\bibfield  {journal} {\bibinfo  {journal} {J. Phys. Chem. Lett.}\ }\textbf {\bibinfo {volume} {14}},\ \bibinfo {pages} {8261} (\bibinfo {year} {2023})}\BibitemShut {NoStop}%
\bibitem [{\citenamefont {Sun}\ and\ \citenamefont {Vendrell}(2023)}]{sun2023modification}%
  \BibitemOpen
  \bibfield  {author} {\bibinfo {author} {\bibfnamefont {J.}~\bibnamefont {Sun}}\ and\ \bibinfo {author} {\bibfnamefont {O.}~\bibnamefont {Vendrell}},\ }\bibfield  {title} {\bibinfo {title} {Modification of thermal chemical rates in a cavity via resonant effects in the collective regime},\ }\href@noop {} {\bibfield  {journal} {\bibinfo  {journal} {J. Phys. Chem. Lett.}\ }\textbf {\bibinfo {volume} {14}},\ \bibinfo {pages} {8397} (\bibinfo {year} {2023})}\BibitemShut {NoStop}%
\bibitem [{\citenamefont {Anderson}\ \emph {et~al.}(2023)\citenamefont {Anderson}, \citenamefont {Woods}, \citenamefont {Fay}, \citenamefont {Wales},\ and\ \citenamefont {Limmer}}]{anderson2023mechanism}%
  \BibitemOpen
  \bibfield  {author} {\bibinfo {author} {\bibfnamefont {M.~C.}\ \bibnamefont {Anderson}}, \bibinfo {author} {\bibfnamefont {E.~J.}\ \bibnamefont {Woods}}, \bibinfo {author} {\bibfnamefont {T.~P.}\ \bibnamefont {Fay}}, \bibinfo {author} {\bibfnamefont {D.~J.}\ \bibnamefont {Wales}},\ and\ \bibinfo {author} {\bibfnamefont {D.~T.}\ \bibnamefont {Limmer}},\ }\bibfield  {title} {\bibinfo {title} {On the mechanism of polaritonic rate suppression from quantum transition paths},\ }\href@noop {} {\bibfield  {journal} {\bibinfo  {journal} {J. Phys. Chem. Lett.}\ }\textbf {\bibinfo {volume} {14}},\ \bibinfo {pages} {6888} (\bibinfo {year} {2023})}\BibitemShut {NoStop}%
\bibitem [{\citenamefont {Sidler}\ \emph {et~al.}(2024{\natexlab{a}})\citenamefont {Sidler}, \citenamefont {Schnappinger}, \citenamefont {Obzhirov}, \citenamefont {Ruggenthaler}, \citenamefont {Kowalewski},\ and\ \citenamefont {Rubio}}]{sidler2024unraveling}%
  \BibitemOpen
  \bibfield  {author} {\bibinfo {author} {\bibfnamefont {D.}~\bibnamefont {Sidler}}, \bibinfo {author} {\bibfnamefont {T.}~\bibnamefont {Schnappinger}}, \bibinfo {author} {\bibfnamefont {A.}~\bibnamefont {Obzhirov}}, \bibinfo {author} {\bibfnamefont {M.}~\bibnamefont {Ruggenthaler}}, \bibinfo {author} {\bibfnamefont {M.}~\bibnamefont {Kowalewski}},\ and\ \bibinfo {author} {\bibfnamefont {A.}~\bibnamefont {Rubio}},\ }\bibfield  {title} {\bibinfo {title} {Unraveling a cavity-induced molecular polarization mechanism from collective vibrational strong coupling},\ }\href@noop {} {\bibfield  {journal} {\bibinfo  {journal} {J. Phys. Chem. Lett.}\ }\textbf {\bibinfo {volume} {15}},\ \bibinfo {pages} {5208} (\bibinfo {year} {2024}{\natexlab{a}})}\BibitemShut {NoStop}%
\bibitem [{\citenamefont {Ying}\ and\ \citenamefont {Huo}(2023)}]{ying2023resonance}%
  \BibitemOpen
  \bibfield  {author} {\bibinfo {author} {\bibfnamefont {W.}~\bibnamefont {Ying}}\ and\ \bibinfo {author} {\bibfnamefont {P.}~\bibnamefont {Huo}},\ }\bibfield  {title} {\bibinfo {title} {Resonance theory and quantum dynamics simulations of vibrational polariton chemistry},\ }\href@noop {} {\bibfield  {journal} {\bibinfo  {journal} {J. of Chem. Phys.}\ }\textbf {\bibinfo {volume} {159}} (\bibinfo {year} {2023})}\BibitemShut {NoStop}%
\bibitem [{\citenamefont {Lindoy}\ \emph {et~al.}(2024)\citenamefont {Lindoy}, \citenamefont {Mandal},\ and\ \citenamefont {Reichman}}]{lindoy2024investigating}%
  \BibitemOpen
  \bibfield  {author} {\bibinfo {author} {\bibfnamefont {L.~P.}\ \bibnamefont {Lindoy}}, \bibinfo {author} {\bibfnamefont {A.}~\bibnamefont {Mandal}},\ and\ \bibinfo {author} {\bibfnamefont {D.~R.}\ \bibnamefont {Reichman}},\ }\bibfield  {title} {\bibinfo {title} {Investigating the collective nature of cavity-modified chemical kinetics under vibrational strong coupling},\ }\href@noop {} {\bibfield  {journal} {\bibinfo  {journal} {Nanophotonics}\ }\textbf {\bibinfo {volume} {13}},\ \bibinfo {pages} {2617} (\bibinfo {year} {2024})}\BibitemShut {NoStop}%
\bibitem [{\citenamefont {Ke}\ and\ \citenamefont {Richardson}(2024)}]{ke2024insights}%
  \BibitemOpen
  \bibfield  {author} {\bibinfo {author} {\bibfnamefont {Y.}~\bibnamefont {Ke}}\ and\ \bibinfo {author} {\bibfnamefont {J.~O.}\ \bibnamefont {Richardson}},\ }\bibfield  {title} {\bibinfo {title} {Insights into the mechanisms of optical cavity-modified ground-state chemical reactions},\ }\href@noop {} {\bibfield  {journal} {\bibinfo  {journal} {J. Chem. Phys.}\ }\textbf {\bibinfo {volume} {160}} (\bibinfo {year} {2024})}\BibitemShut {NoStop}%
\bibitem [{\citenamefont {Sidler}\ \emph {et~al.}(2024{\natexlab{b}})\citenamefont {Sidler}, \citenamefont {Ruggenthaler},\ and\ \citenamefont {Rubio}}]{sidler2024connection}%
  \BibitemOpen
  \bibfield  {author} {\bibinfo {author} {\bibfnamefont {D.}~\bibnamefont {Sidler}}, \bibinfo {author} {\bibfnamefont {M.}~\bibnamefont {Ruggenthaler}},\ and\ \bibinfo {author} {\bibfnamefont {A.}~\bibnamefont {Rubio}},\ }\bibfield  {title} {\bibinfo {title} {The connection of polaritonic chemistry with the physics of a spin glass},\ }\href@noop {} {\bibfield  {journal} {\bibinfo  {journal} {arXiv preprint arXiv:2409.08986}\ } (\bibinfo {year} {2024}{\natexlab{b}})}\BibitemShut {NoStop}%
\bibitem [{\citenamefont {Ying}\ \emph {et~al.}(2024)\citenamefont {Ying}, \citenamefont {Taylor},\ and\ \citenamefont {Huo}}]{ying2024resonance}%
  \BibitemOpen
  \bibfield  {author} {\bibinfo {author} {\bibfnamefont {W.}~\bibnamefont {Ying}}, \bibinfo {author} {\bibfnamefont {M.~A.}\ \bibnamefont {Taylor}},\ and\ \bibinfo {author} {\bibfnamefont {P.}~\bibnamefont {Huo}},\ }\bibfield  {title} {\bibinfo {title} {Resonance theory of vibrational polariton chemistry at the normal incidence},\ }\href@noop {} {\bibfield  {journal} {\bibinfo  {journal} {Nanophotonics}\ }\textbf {\bibinfo {volume} {13}},\ \bibinfo {pages} {2601} (\bibinfo {year} {2024})}\BibitemShut {NoStop}%
\bibitem [{\citenamefont {Ke}(2025)}]{ke2025stochastic}%
  \BibitemOpen
  \bibfield  {author} {\bibinfo {author} {\bibfnamefont {Y.}~\bibnamefont {Ke}},\ }\bibfield  {title} {\bibinfo {title} {Stochastic resonance in vibrational polariton chemistry},\ }\href@noop {} {\bibfield  {journal} {\bibinfo  {journal} {J. Chem. Phys.}\ }\textbf {\bibinfo {volume} {162}} (\bibinfo {year} {2025})}\BibitemShut {NoStop}%
\bibitem [{\citenamefont {Horak}\ \emph {et~al.}(2025)\citenamefont {Horak}, \citenamefont {Sidler}, \citenamefont {Schnappinger}, \citenamefont {Huang}, \citenamefont {Ruggenthaler},\ and\ \citenamefont {Rubio}}]{horak2025analytic}%
  \BibitemOpen
  \bibfield  {author} {\bibinfo {author} {\bibfnamefont {J.}~\bibnamefont {Horak}}, \bibinfo {author} {\bibfnamefont {D.}~\bibnamefont {Sidler}}, \bibinfo {author} {\bibfnamefont {T.}~\bibnamefont {Schnappinger}}, \bibinfo {author} {\bibfnamefont {W.-M.}\ \bibnamefont {Huang}}, \bibinfo {author} {\bibfnamefont {M.}~\bibnamefont {Ruggenthaler}},\ and\ \bibinfo {author} {\bibfnamefont {A.}~\bibnamefont {Rubio}},\ }\bibfield  {title} {\bibinfo {title} {Analytic model reveals local molecular polarizability changes induced by collective strong coupling in optical cavities},\ }\href@noop {} {\bibfield  {journal} {\bibinfo  {journal} {Phys. Rev. Res.}\ }\textbf {\bibinfo {volume} {7}},\ \bibinfo {pages} {013242} (\bibinfo {year} {2025})}\BibitemShut {NoStop}%
\bibitem [{\citenamefont {Vega}\ \emph {et~al.}(2025)\citenamefont {Vega}, \citenamefont {Ying},\ and\ \citenamefont {Huo}}]{vega2025theoretical}%
  \BibitemOpen
  \bibfield  {author} {\bibinfo {author} {\bibfnamefont {S.~M.}\ \bibnamefont {Vega}}, \bibinfo {author} {\bibfnamefont {W.}~\bibnamefont {Ying}},\ and\ \bibinfo {author} {\bibfnamefont {P.}~\bibnamefont {Huo}},\ }\bibfield  {title} {\bibinfo {title} {Theoretical insights into the resonant suppression effect in vibrational polariton chemistry},\ }\href@noop {} {\bibfield  {journal} {\bibinfo  {journal} {J. Am. Chem. Soc.}\ } (\bibinfo {year} {2025})}\BibitemShut {NoStop}%
\bibitem [{\citenamefont {Rokaj}\ \emph {et~al.}(2018)\citenamefont {Rokaj}, \citenamefont {Welakuh}, \citenamefont {Ruggenthaler},\ and\ \citenamefont {Rubio}}]{rokaj2018light}%
  \BibitemOpen
  \bibfield  {author} {\bibinfo {author} {\bibfnamefont {V.}~\bibnamefont {Rokaj}}, \bibinfo {author} {\bibfnamefont {D.~M.}\ \bibnamefont {Welakuh}}, \bibinfo {author} {\bibfnamefont {M.}~\bibnamefont {Ruggenthaler}},\ and\ \bibinfo {author} {\bibfnamefont {A.}~\bibnamefont {Rubio}},\ }\bibfield  {title} {\bibinfo {title} {Light--matter interaction in the long-wavelength limit: no ground-state without dipole self-energy},\ }\href@noop {} {\bibfield  {journal} {\bibinfo  {journal} {J. Phys. B -- At. Mol. Opt.}\ }\textbf {\bibinfo {volume} {51}},\ \bibinfo {pages} {034005} (\bibinfo {year} {2018})}\BibitemShut {NoStop}%
\bibitem [{\citenamefont {Sch\"{a}fer}\ \emph {et~al.}(2020)\citenamefont {Sch\"{a}fer}, \citenamefont {Ruggenthaler}, \citenamefont {Rokaj},\ and\ \citenamefont {Rubio}}]{schafer2020relevance}%
  \BibitemOpen
  \bibfield  {author} {\bibinfo {author} {\bibfnamefont {C.}~\bibnamefont {Sch\"{a}fer}}, \bibinfo {author} {\bibfnamefont {M.}~\bibnamefont {Ruggenthaler}}, \bibinfo {author} {\bibfnamefont {V.}~\bibnamefont {Rokaj}},\ and\ \bibinfo {author} {\bibfnamefont {A.}~\bibnamefont {Rubio}},\ }\bibfield  {title} {\bibinfo {title} {Relevance of the quadratic diamagnetic and self-polarization terms in cavity quantum electrodynamics},\ }\href@noop {} {\bibfield  {journal} {\bibinfo  {journal} {ACS Photonics}\ }\textbf {\bibinfo {volume} {7}},\ \bibinfo {pages} {975} (\bibinfo {year} {2020})}\BibitemShut {NoStop}%
\bibitem [{\citenamefont {Fregoni}\ \emph {et~al.}(2022)\citenamefont {Fregoni}, \citenamefont {Garcia-Vidal},\ and\ \citenamefont {Feist}}]{fregoni2022theoretical}%
  \BibitemOpen
  \bibfield  {author} {\bibinfo {author} {\bibfnamefont {J.}~\bibnamefont {Fregoni}}, \bibinfo {author} {\bibfnamefont {F.~J.}\ \bibnamefont {Garcia-Vidal}},\ and\ \bibinfo {author} {\bibfnamefont {J.}~\bibnamefont {Feist}},\ }\bibfield  {title} {\bibinfo {title} {Theoretical challenges in polaritonic chemistry},\ }\href@noop {} {\bibfield  {journal} {\bibinfo  {journal} {ACS Photonics}\ }\textbf {\bibinfo {volume} {9}},\ \bibinfo {pages} {1096} (\bibinfo {year} {2022})}\BibitemShut {NoStop}%
\bibitem [{\citenamefont {de~la Pradilla}\ \emph {et~al.}(2025)\citenamefont {de~la Pradilla}, \citenamefont {Moreno},\ and\ \citenamefont {Feist}}]{delapradilla2025there}%
  \BibitemOpen
  \bibfield  {author} {\bibinfo {author} {\bibfnamefont {D.~F.}\ \bibnamefont {de~la Pradilla}}, \bibinfo {author} {\bibfnamefont {E.}~\bibnamefont {Moreno}},\ and\ \bibinfo {author} {\bibfnamefont {J.}~\bibnamefont {Feist}},\ }\bibfield  {title} {\bibinfo {title} {There is no ultrastrong coupling with photons},\ }\href@noop {} {\bibfield  {journal} {\bibinfo  {journal} {arXiv preprint arXiv:2508.00702}\ } (\bibinfo {year} {2025})}\BibitemShut {NoStop}%
\bibitem [{\citenamefont {Mandal}\ \emph {et~al.}(2023{\natexlab{b}})\citenamefont {Mandal}, \citenamefont {Xu}, \citenamefont {Mahajan}, \citenamefont {Lee}, \citenamefont {Delor},\ and\ \citenamefont {Reichman}}]{mandal2023microscopic}%
  \BibitemOpen
  \bibfield  {author} {\bibinfo {author} {\bibfnamefont {A.}~\bibnamefont {Mandal}}, \bibinfo {author} {\bibfnamefont {D.}~\bibnamefont {Xu}}, \bibinfo {author} {\bibfnamefont {A.}~\bibnamefont {Mahajan}}, \bibinfo {author} {\bibfnamefont {J.}~\bibnamefont {Lee}}, \bibinfo {author} {\bibfnamefont {M.}~\bibnamefont {Delor}},\ and\ \bibinfo {author} {\bibfnamefont {D.~R.}\ \bibnamefont {Reichman}},\ }\bibfield  {title} {\bibinfo {title} {Microscopic theory of multimode polariton dispersion in multilayered materials},\ }\href@noop {} {\bibfield  {journal} {\bibinfo  {journal} {Nano Lett.}\ }\textbf {\bibinfo {volume} {23}},\ \bibinfo {pages} {4082} (\bibinfo {year} {2023}{\natexlab{b}})}\BibitemShut {NoStop}%
\bibitem [{\citenamefont {Fiechter}\ and\ \citenamefont {Richardson}(2024)}]{fiechter2024understanding}%
  \BibitemOpen
  \bibfield  {author} {\bibinfo {author} {\bibfnamefont {M.~R.}\ \bibnamefont {Fiechter}}\ and\ \bibinfo {author} {\bibfnamefont {J.~O.}\ \bibnamefont {Richardson}},\ }\bibfield  {title} {\bibinfo {title} {Understanding the cavity {Born--Oppenheimer} approximation},\ }\href@noop {} {\bibfield  {journal} {\bibinfo  {journal} {J. Chem. Phys.}\ }\textbf {\bibinfo {volume} {160}} (\bibinfo {year} {2024})}\BibitemShut {NoStop}%
\bibitem [{\citenamefont {Schnappinger}\ \emph {et~al.}(2023)\citenamefont {Schnappinger}, \citenamefont {Sidler}, \citenamefont {Ruggenthaler}, \citenamefont {Rubio},\ and\ \citenamefont {Kowalewski}}]{schnappinger2023PES}%
  \BibitemOpen
  \bibfield  {author} {\bibinfo {author} {\bibfnamefont {T.}~\bibnamefont {Schnappinger}}, \bibinfo {author} {\bibfnamefont {D.}~\bibnamefont {Sidler}}, \bibinfo {author} {\bibfnamefont {M.}~\bibnamefont {Ruggenthaler}}, \bibinfo {author} {\bibfnamefont {A.}~\bibnamefont {Rubio}},\ and\ \bibinfo {author} {\bibfnamefont {M.}~\bibnamefont {Kowalewski}},\ }\bibfield  {title} {\bibinfo {title} {Cavity {Born--Oppenheimer Hartree--Fock Ansatz}: Light--matter properties of strongly coupled molecular ensembles},\ }\href@noop {} {\bibfield  {journal} {\bibinfo  {journal} {J. Phys. Chem. Lett.}\ }\textbf {\bibinfo {volume} {14}},\ \bibinfo {pages} {8024} (\bibinfo {year} {2023})}\BibitemShut {NoStop}%
\bibitem [{\citenamefont {Haugland}\ \emph {et~al.}(2025)\citenamefont {Haugland}, \citenamefont {Philbin}, \citenamefont {Ghosh}, \citenamefont {Chen}, \citenamefont {Koch},\ and\ \citenamefont {Narang}}]{haugland2023understanding}%
  \BibitemOpen
  \bibfield  {author} {\bibinfo {author} {\bibfnamefont {T.~S.}\ \bibnamefont {Haugland}}, \bibinfo {author} {\bibfnamefont {J.~P.}\ \bibnamefont {Philbin}}, \bibinfo {author} {\bibfnamefont {T.~K.}\ \bibnamefont {Ghosh}}, \bibinfo {author} {\bibfnamefont {M.}~\bibnamefont {Chen}}, \bibinfo {author} {\bibfnamefont {H.}~\bibnamefont {Koch}},\ and\ \bibinfo {author} {\bibfnamefont {P.}~\bibnamefont {Narang}},\ }\bibfield  {title} {\bibinfo {title} {Understanding the polaritonic ground state in cavity quantum electrodynamics},\ }\href@noop {} {\bibfield  {journal} {\bibinfo  {journal} {J. Chem. Phys.}\ }\textbf {\bibinfo {volume} {162}} (\bibinfo {year} {2025})}\BibitemShut {NoStop}%
\bibitem [{\citenamefont {Fischer}(2024)}]{fischer2024cavity}%
  \BibitemOpen
  \bibfield  {author} {\bibinfo {author} {\bibfnamefont {E.~W.}\ \bibnamefont {Fischer}},\ }\bibfield  {title} {\bibinfo {title} {Cavity-modified local and non-local electronic interactions in molecular ensembles under vibrational strong coupling},\ }\href@noop {} {\bibfield  {journal} {\bibinfo  {journal} {J. Chem. Phys.}\ }\textbf {\bibinfo {volume} {161}} (\bibinfo {year} {2024})}\BibitemShut {NoStop}%
\bibitem [{\citenamefont {Borges}\ \emph {et~al.}(2024)\citenamefont {Borges}, \citenamefont {Schnappinger},\ and\ \citenamefont {Kowalewski}}]{borges2024extending}%
  \BibitemOpen
  \bibfield  {author} {\bibinfo {author} {\bibfnamefont {L.}~\bibnamefont {Borges}}, \bibinfo {author} {\bibfnamefont {T.}~\bibnamefont {Schnappinger}},\ and\ \bibinfo {author} {\bibfnamefont {M.}~\bibnamefont {Kowalewski}},\ }\bibfield  {title} {\bibinfo {title} {Extending the {Tavis--Cummings} model for molecular ensembles—exploring the effects of dipole self-energies and static dipole moments},\ }\href@noop {} {\bibfield  {journal} {\bibinfo  {journal} {J. Chem. Phys.}\ }\textbf {\bibinfo {volume} {161}} (\bibinfo {year} {2024})}\BibitemShut {NoStop}%
\bibitem [{\citenamefont {Cohen-Tannoudji}\ \emph {et~al.}(1997)\citenamefont {Cohen-Tannoudji}, \citenamefont {Dupont-Roc},\ and\ \citenamefont {Grynberg}}]{cohen1997photons}%
  \BibitemOpen
  \bibfield  {author} {\bibinfo {author} {\bibfnamefont {C.}~\bibnamefont {Cohen-Tannoudji}}, \bibinfo {author} {\bibfnamefont {J.}~\bibnamefont {Dupont-Roc}},\ and\ \bibinfo {author} {\bibfnamefont {G.}~\bibnamefont {Grynberg}},\ }\href@noop {} {\emph {\bibinfo {title} {Photons and atoms -- introduction to quantum electrodynamics}}}\ (\bibinfo  {publisher} {John Wiley \& Sons, Ltd},\ \bibinfo {year} {1997})\BibitemShut {NoStop}%
\bibitem [{\citenamefont {Keeling}()}]{keelingQO}%
  \BibitemOpen
  \bibfield  {author} {\bibinfo {author} {\bibfnamefont {J.}~\bibnamefont {Keeling}},\ }\href@noop {} {\bibinfo {title} {Lecture notes: Light-matter interactions and quantum optics.}}\BibitemShut {Stop}%
\bibitem [{\citenamefont {Power}\ and\ \citenamefont {Thirunamachandran}(1982)}]{power1982quantum}%
  \BibitemOpen
  \bibfield  {author} {\bibinfo {author} {\bibfnamefont {E.}~\bibnamefont {Power}}\ and\ \bibinfo {author} {\bibfnamefont {T.}~\bibnamefont {Thirunamachandran}},\ }\bibfield  {title} {\bibinfo {title} {Quantum electrodynamics in a cavity},\ }\href@noop {} {\bibfield  {journal} {\bibinfo  {journal} {Phys. Rev. A}\ }\textbf {\bibinfo {volume} {25}},\ \bibinfo {pages} {2473} (\bibinfo {year} {1982})}\BibitemShut {NoStop}%
\bibitem [{\citenamefont {Vukics}\ and\ \citenamefont {Domokos}(2012)}]{vukics2012adequacy}%
  \BibitemOpen
  \bibfield  {author} {\bibinfo {author} {\bibfnamefont {A.}~\bibnamefont {Vukics}}\ and\ \bibinfo {author} {\bibfnamefont {P.}~\bibnamefont {Domokos}},\ }\bibfield  {title} {\bibinfo {title} {Adequacy of the {D}icke model in cavity {QED}: A counter-no-go statement},\ }\href@noop {} {\bibfield  {journal} {\bibinfo  {journal} {Phys. Rev. A}\ }\textbf {\bibinfo {volume} {86}},\ \bibinfo {pages} {053807} (\bibinfo {year} {2012})}\BibitemShut {NoStop}%
\bibitem [{\citenamefont {De~Bernardis}\ \emph {et~al.}(2018)\citenamefont {De~Bernardis}, \citenamefont {Jaako},\ and\ \citenamefont {Rabl}}]{de2018cavity}%
  \BibitemOpen
  \bibfield  {author} {\bibinfo {author} {\bibfnamefont {D.}~\bibnamefont {De~Bernardis}}, \bibinfo {author} {\bibfnamefont {T.}~\bibnamefont {Jaako}},\ and\ \bibinfo {author} {\bibfnamefont {P.}~\bibnamefont {Rabl}},\ }\bibfield  {title} {\bibinfo {title} {Cavity quantum electrodynamics in the nonperturbative regime},\ }\href@noop {} {\bibfield  {journal} {\bibinfo  {journal} {Phys. Rev. A}\ }\textbf {\bibinfo {volume} {97}},\ \bibinfo {pages} {043820} (\bibinfo {year} {2018})}\BibitemShut {NoStop}%
\bibitem [{\citenamefont {Buhmann}(2013)}]{buhmann2013dispersion_book}%
  \BibitemOpen
  \bibfield  {author} {\bibinfo {author} {\bibfnamefont {S.~Y.}\ \bibnamefont {Buhmann}},\ }\href@noop {} {\emph {\bibinfo {title} {Dispersion Forces I: Macroscopic quantum electrodynamics and ground-state {Casimir, Casimir--Polder and van der Waals forces}}}},\ Vol.\ \bibinfo {volume} {247}\ (\bibinfo  {publisher} {Springer},\ \bibinfo {year} {2013})\BibitemShut {NoStop}%
\bibitem [{Note1()}]{Note1}%
  \BibitemOpen
  \bibinfo {note} {Note that by doing this multi-center dipole transformation, we break the symmetry of the Hamiltonian under exchange of particles (second line of Eq.~\protect \eqref {eq:fullHfree}): we effectively distinguish the charged particles (\protect \emph {i.e.} electrons/nuclei) that make up dipole $\protect \hat {\protect \bm {\mu }}_A$ from those in dipole $\protect \hat {\protect \bm {\mu }}_B$ by assigning the former to be close to $\protect \mathbf {r}_A$, and the latter to $\protect \mathbf {r}_B$. In the context of electronic structure calculations, this means that we neglect the exchange interaction between electrons in different molecules; as this interaction decays exponentially, it is expected to be negligible if $A$ and $B$ are separated by a distance multiple times their size \cite {stone2013theory}, as assumed in the text.}\BibitemShut {Stop}%
\bibitem [{Note2()}]{Note2}%
  \BibitemOpen
  \bibinfo {note} {Or more precisely, dipole moment matrix elements -- \protect \emph {i.e.}, transition dipole moments as well as permanent dipole moments.}\BibitemShut {Stop}%
\bibitem [{Note3()}]{Note3}%
  \BibitemOpen
  \bibinfo {note} {By increasing the value of $Q$, we can decrease the spatial extent of the dipoles, and eventually turn them into point dipoles; this guarantees us that their interaction is purely a dipole--dipole interaction.}\BibitemShut {Stop}%
\bibitem [{\citenamefont {Salam}(2009)}]{salam2009molecular}%
  \BibitemOpen
  \bibfield  {author} {\bibinfo {author} {\bibfnamefont {A.}~\bibnamefont {Salam}},\ }\href@noop {} {\emph {\bibinfo {title} {Molecular quantum electrodynamics: long-range intermolecular interactions}}}\ (\bibinfo  {publisher} {John Wiley \& Sons},\ \bibinfo {year} {2009})\BibitemShut {NoStop}%
\bibitem [{\citenamefont {Woolley}(2024)}]{woolley2024infinities}%
  \BibitemOpen
  \bibfield  {author} {\bibinfo {author} {\bibfnamefont {R.~G.}\ \bibnamefont {Woolley}},\ }\bibfield  {title} {\bibinfo {title} {Infinities in molecular quantum electrodynamics and generalized functions},\ }\href@noop {} {\bibfield  {journal} {\bibinfo  {journal} {Phys. Rev. A}\ }\textbf {\bibinfo {volume} {110}},\ \bibinfo {pages} {012204} (\bibinfo {year} {2024})}\BibitemShut {NoStop}%
\bibitem [{\citenamefont {Schwinger}\ \emph {et~al.}(1998)\citenamefont {Schwinger}, \citenamefont {DeRaad~Jr}, \citenamefont {Milton},\ and\ \citenamefont {Tsai}}]{schwinger2019classical}%
  \BibitemOpen
  \bibfield  {author} {\bibinfo {author} {\bibfnamefont {J.}~\bibnamefont {Schwinger}}, \bibinfo {author} {\bibfnamefont {L.~L.}\ \bibnamefont {DeRaad~Jr}}, \bibinfo {author} {\bibfnamefont {K.~A.}\ \bibnamefont {Milton}},\ and\ \bibinfo {author} {\bibfnamefont {W.-y.}\ \bibnamefont {Tsai}},\ }\href@noop {} {\emph {\bibinfo {title} {Classical Electrodynamics}}}\ (\bibinfo  {publisher} {Perseus Books},\ \bibinfo {year} {1998})\BibitemShut {NoStop}%
\bibitem [{\citenamefont {Einziger}\ \emph {et~al.}(2002)\citenamefont {Einziger}, \citenamefont {Livshitz},\ and\ \citenamefont {Mizrahi}}]{einziger2002rigorous}%
  \BibitemOpen
  \bibfield  {author} {\bibinfo {author} {\bibfnamefont {P.~D.}\ \bibnamefont {Einziger}}, \bibinfo {author} {\bibfnamefont {L.~M.}\ \bibnamefont {Livshitz}},\ and\ \bibinfo {author} {\bibfnamefont {J.}~\bibnamefont {Mizrahi}},\ }\bibfield  {title} {\bibinfo {title} {Rigorous image-series expansions of quasi-static {Green}'s functions for regions with planar stratification},\ }\href@noop {} {\bibfield  {journal} {\bibinfo  {journal} {IEEE T. Antenn. Propag.}\ }\textbf {\bibinfo {volume} {50}},\ \bibinfo {pages} {1813} (\bibinfo {year} {2002})}\BibitemShut {NoStop}%
\bibitem [{\citenamefont {Barcellona}\ \emph {et~al.}(2018)\citenamefont {Barcellona}, \citenamefont {Bennett},\ and\ \citenamefont {Buhmann}}]{Barcellona_2018}%
  \BibitemOpen
  \bibfield  {author} {\bibinfo {author} {\bibfnamefont {P.}~\bibnamefont {Barcellona}}, \bibinfo {author} {\bibfnamefont {R.}~\bibnamefont {Bennett}},\ and\ \bibinfo {author} {\bibfnamefont {S.~Y.}\ \bibnamefont {Buhmann}},\ }\bibfield  {title} {\bibinfo {title} {Manipulating the {Coulomb} interaction: {A} {Green}'s function perspective},\ }\href {https://doi.org/10.1088/2399-6528/aaa70a} {\bibfield  {journal} {\bibinfo  {journal} {J. Phys. Commun.}\ }\textbf {\bibinfo {volume} {2}},\ \bibinfo {pages} {035027} (\bibinfo {year} {2018})}\BibitemShut {NoStop}%
\bibitem [{Note4()}]{Note4}%
  \BibitemOpen
  \bibinfo {note} {The method of images is a well-known approach in classical electrodynamics, but can also be derived in the framework of quantum electrodynamics, where it is associated with the exchange of a virtual photon between a charge and an image charge \cite {Barcellona_2018}. Do note that this latter derivation makes use of perturbation theory.}\BibitemShut {Stop}%
\bibitem [{\citenamefont {Griffiths}(2023)}]{griffiths2023ED}%
  \BibitemOpen
  \bibfield  {author} {\bibinfo {author} {\bibfnamefont {D.~J.}\ \bibnamefont {Griffiths}},\ }\href@noop {} {\emph {\bibinfo {title} {Introduction to electrodynamics}}}\ (\bibinfo  {publisher} {Cambridge University Press},\ \bibinfo {year} {2023})\BibitemShut {NoStop}%
\bibitem [{\citenamefont {Meschede}\ \emph {et~al.}(1990)\citenamefont {Meschede}, \citenamefont {Jhe},\ and\ \citenamefont {Hinds}}]{meschede1990radiative}%
  \BibitemOpen
  \bibfield  {author} {\bibinfo {author} {\bibfnamefont {D.}~\bibnamefont {Meschede}}, \bibinfo {author} {\bibfnamefont {W.}~\bibnamefont {Jhe}},\ and\ \bibinfo {author} {\bibfnamefont {E.}~\bibnamefont {Hinds}},\ }\bibfield  {title} {\bibinfo {title} {Radiative properties of atoms near a conducting plane: An old problem in a new light},\ }\href@noop {} {\bibfield  {journal} {\bibinfo  {journal} {Phys. Rev. A}\ }\textbf {\bibinfo {volume} {41}},\ \bibinfo {pages} {1587} (\bibinfo {year} {1990})}\BibitemShut {NoStop}%
\bibitem [{Note5()}]{Note5}%
  \BibitemOpen
  \bibinfo {note} {Note that in contrast to Vukics \protect \emph {et al.} \cite {vukics2012adequacy}, we let the sum over $k_z$ run over negative as well as positive values; our normalization $A_\protect \mathbf {k}$ therefore deviates from theirs accordingly}\BibitemShut {NoStop}%
\bibitem [{Note6()}]{Note6}%
  \BibitemOpen
  \bibinfo {note} {We used here that we can flip the signs of $k_z$ and $\protect \mathbf {k}_\parallel $ under the integral. Flipping the sign of $k_z$ has the effect $\protect \mathbf {e}_\protect \mathrm {p}\rightarrow \protect \mathbf {R}_{xy}\protect \mathbf {e}_\protect \mathrm {p}$ while leaving $\protect \mathbf {e}_\protect \mathrm {s}$ unchanged. We have also used that $\protect \mathbf {R}_{xy}\protect \mathbf {e}_\protect \mathrm {s}=-\protect \mathbf {e}_\protect \mathrm {s}$. These properties can easily be verified using the explicit form of the polarization vectors given in the SI}\BibitemShut {NoStop}%
\bibitem [{\citenamefont {Ley}\ and\ \citenamefont {Loudon}(1987)}]{ley1987quantum}%
  \BibitemOpen
  \bibfield  {author} {\bibinfo {author} {\bibfnamefont {M.}~\bibnamefont {Ley}}\ and\ \bibinfo {author} {\bibfnamefont {R.}~\bibnamefont {Loudon}},\ }\bibfield  {title} {\bibinfo {title} {Quantum theory of high-resolution length measurement with a {Fabry--Perot} interferometer},\ }\href@noop {} {\bibfield  {journal} {\bibinfo  {journal} {J. Mod. Optic.}\ }\textbf {\bibinfo {volume} {34}},\ \bibinfo {pages} {227} (\bibinfo {year} {1987})}\BibitemShut {NoStop}%
\bibitem [{\citenamefont {De~Martini}\ \emph {et~al.}(1991)\citenamefont {De~Martini}, \citenamefont {Marrocco}, \citenamefont {Mataloni}, \citenamefont {Crescentini},\ and\ \citenamefont {Loudon}}]{de1991spontaneous}%
  \BibitemOpen
  \bibfield  {author} {\bibinfo {author} {\bibfnamefont {F.}~\bibnamefont {De~Martini}}, \bibinfo {author} {\bibfnamefont {M.}~\bibnamefont {Marrocco}}, \bibinfo {author} {\bibfnamefont {P.}~\bibnamefont {Mataloni}}, \bibinfo {author} {\bibfnamefont {L.}~\bibnamefont {Crescentini}},\ and\ \bibinfo {author} {\bibfnamefont {R.}~\bibnamefont {Loudon}},\ }\bibfield  {title} {\bibinfo {title} {Spontaneous emission in the optical microscopic cavity},\ }\href@noop {} {\bibfield  {journal} {\bibinfo  {journal} {Phys. Rev. A}\ }\textbf {\bibinfo {volume} {43}},\ \bibinfo {pages} {2480} (\bibinfo {year} {1991})}\BibitemShut {NoStop}%
\bibitem [{\citenamefont {Svendsen}\ \emph {et~al.}(2023)\citenamefont {Svendsen}, \citenamefont {Ruggenthaler}, \citenamefont {H{\"u}bener}, \citenamefont {Sch{\"a}fer}, \citenamefont {Eckstein}, \citenamefont {Rubio},\ and\ \citenamefont {Latini}}]{svendsen2023theory}%
  \BibitemOpen
  \bibfield  {author} {\bibinfo {author} {\bibfnamefont {M.~K.}\ \bibnamefont {Svendsen}}, \bibinfo {author} {\bibfnamefont {M.}~\bibnamefont {Ruggenthaler}}, \bibinfo {author} {\bibfnamefont {H.}~\bibnamefont {H{\"u}bener}}, \bibinfo {author} {\bibfnamefont {C.}~\bibnamefont {Sch{\"a}fer}}, \bibinfo {author} {\bibfnamefont {M.}~\bibnamefont {Eckstein}}, \bibinfo {author} {\bibfnamefont {A.}~\bibnamefont {Rubio}},\ and\ \bibinfo {author} {\bibfnamefont {S.}~\bibnamefont {Latini}},\ }\bibfield  {title} {\bibinfo {title} {Theory of quantum light-matter interaction in cavities: Extended systems and the long wavelength approximation},\ }\href@noop {} {\bibfield  {journal} {\bibinfo  {journal} {arXiv preprint arXiv:2312.17374}\ } (\bibinfo {year} {2023})}\BibitemShut {NoStop}%
\bibitem [{\citenamefont {Dutra}\ and\ \citenamefont {Knight}(1996)}]{dutra1996spontaneous}%
  \BibitemOpen
  \bibfield  {author} {\bibinfo {author} {\bibfnamefont {S.}~\bibnamefont {Dutra}}\ and\ \bibinfo {author} {\bibfnamefont {P.}~\bibnamefont {Knight}},\ }\bibfield  {title} {\bibinfo {title} {Spontaneous emission in a planar {Fabry--P{\'e}rot} microcavity},\ }\href@noop {} {\bibfield  {journal} {\bibinfo  {journal} {Phys. Rev. A}\ }\textbf {\bibinfo {volume} {53}},\ \bibinfo {pages} {3587} (\bibinfo {year} {1996})}\BibitemShut {NoStop}%
\bibitem [{\citenamefont {Suter}\ and\ \citenamefont {Dietiker}(2014)}]{suter2014calculation}%
  \BibitemOpen
  \bibfield  {author} {\bibinfo {author} {\bibfnamefont {M.}~\bibnamefont {Suter}}\ and\ \bibinfo {author} {\bibfnamefont {P.}~\bibnamefont {Dietiker}},\ }\bibfield  {title} {\bibinfo {title} {Calculation of the finesse of an ideal {Fabry--Perot }resonator},\ }\href@noop {} {\bibfield  {journal} {\bibinfo  {journal} {Appl. Opt.}\ }\textbf {\bibinfo {volume} {53}},\ \bibinfo {pages} {7004} (\bibinfo {year} {2014})}\BibitemShut {NoStop}%
\bibitem [{Note7()}]{Note7}%
  \BibitemOpen
  \bibinfo {note} {By this we mean that we effectively calculate the expectation values of the DSE and Coulombic dipole--dipole interaction for a given state. For simplicity, we use a product state, $\ket {\psi }=\ket {\phi }_A\ket {\phi }_B$ for this, so that we can just replace the operators $\protect \hat {\protect \bm {\mu }}_i$ by their expectation value $\bra {\phi }_i \protect \,\protect \hat {\protect \bm {\mu }}_i \ket {\phi }_i$. It is however straightforward to generalize this to a linear combination of states of molecules $A$ and $B$.}\BibitemShut {Stop}%
\bibitem [{tho()}]{thorlabsgold}%
  \BibitemOpen
  \href@noop {} {\emph {\bibinfo {title} {Gold Coating Reflectance, 45° AOI}}},\ \bibinfo {organization} {Thorlabs},\ \bibinfo {note} {available at \url{https://www.thorlabs.com/newgrouppage9.cfm?objectgroup_ID=744}}\BibitemShut {NoStop}%
\bibitem [{\citenamefont {Zangwill}(2013)}]{zangwill2013modern}%
  \BibitemOpen
  \bibfield  {author} {\bibinfo {author} {\bibfnamefont {A.}~\bibnamefont {Zangwill}},\ }\href@noop {} {\emph {\bibinfo {title} {Modern electrodynamics}}}\ (\bibinfo  {publisher} {Cambridge University Press},\ \bibinfo {year} {2013})\BibitemShut {NoStop}%
\bibitem [{\citenamefont {Scheel}\ and\ \citenamefont {Buhmann}(2009)}]{scheel2009macroscopic}%
  \BibitemOpen
  \bibfield  {author} {\bibinfo {author} {\bibfnamefont {S.}~\bibnamefont {Scheel}}\ and\ \bibinfo {author} {\bibfnamefont {S.~Y.}\ \bibnamefont {Buhmann}},\ }\bibfield  {title} {\bibinfo {title} {Macroscopic {QED} -- concepts and applications},\ }\href@noop {} {\bibfield  {journal} {\bibinfo  {journal} {arXiv preprint arXiv:0902.3586}\ } (\bibinfo {year} {2009})}\BibitemShut {NoStop}%
\bibitem [{\citenamefont {Feist}\ \emph {et~al.}(2020)\citenamefont {Feist}, \citenamefont {Fern{\'a}ndez-Dom{\'\i}nguez},\ and\ \citenamefont {Garc{\'\i}a-Vidal}}]{feist2020macroscopic}%
  \BibitemOpen
  \bibfield  {author} {\bibinfo {author} {\bibfnamefont {J.}~\bibnamefont {Feist}}, \bibinfo {author} {\bibfnamefont {A.~I.}\ \bibnamefont {Fern{\'a}ndez-Dom{\'\i}nguez}},\ and\ \bibinfo {author} {\bibfnamefont {F.~J.}\ \bibnamefont {Garc{\'\i}a-Vidal}},\ }\bibfield  {title} {\bibinfo {title} {Macroscopic {QED} for quantum nanophotonics: emitter-centered modes as a minimal basis for multiemitter problems},\ }\href@noop {} {\bibfield  {journal} {\bibinfo  {journal} {Nanophotonics}\ }\textbf {\bibinfo {volume} {10}},\ \bibinfo {pages} {477} (\bibinfo {year} {2020})}\BibitemShut {NoStop}%
\bibitem [{\citenamefont {Hsu}(2025)}]{hsu2025chemistry}%
  \BibitemOpen
  \bibfield  {author} {\bibinfo {author} {\bibfnamefont {L.-Y.}\ \bibnamefont {Hsu}},\ }\bibfield  {title} {\bibinfo {title} {Chemistry meets plasmon polaritons and cavity photons: A perspective from macroscopic quantum electrodynamics},\ }\href@noop {} {\bibfield  {journal} {\bibinfo  {journal} {J. Phys. Chem. Lett.}\ }\textbf {\bibinfo {volume} {16}},\ \bibinfo {pages} {1604} (\bibinfo {year} {2025})}\BibitemShut {NoStop}%
\bibitem [{\citenamefont {Huttner}\ and\ \citenamefont {Barnett}(1992{\natexlab{a}})}]{huttner1992quantization}%
  \BibitemOpen
  \bibfield  {author} {\bibinfo {author} {\bibfnamefont {B.}~\bibnamefont {Huttner}}\ and\ \bibinfo {author} {\bibfnamefont {S.~M.}\ \bibnamefont {Barnett}},\ }\bibfield  {title} {\bibinfo {title} {Quantization of the electromagnetic field in dielectrics},\ }\href@noop {} {\bibfield  {journal} {\bibinfo  {journal} {Phys. Rev. A}\ }\textbf {\bibinfo {volume} {46}},\ \bibinfo {pages} {4306} (\bibinfo {year} {1992}{\natexlab{a}})}\BibitemShut {NoStop}%
\bibitem [{\citenamefont {Huttner}\ and\ \citenamefont {Barnett}(1992{\natexlab{b}})}]{huttner1992dispersion}%
  \BibitemOpen
  \bibfield  {author} {\bibinfo {author} {\bibfnamefont {B.}~\bibnamefont {Huttner}}\ and\ \bibinfo {author} {\bibfnamefont {S.}~\bibnamefont {Barnett}},\ }\bibfield  {title} {\bibinfo {title} {Dispersion and loss in a {Hopfield} dielectric},\ }\href@noop {} {\bibfield  {journal} {\bibinfo  {journal} {Europhys. Lett.}\ }\textbf {\bibinfo {volume} {18}},\ \bibinfo {pages} {487} (\bibinfo {year} {1992}{\natexlab{b}})}\BibitemShut {NoStop}%
\bibitem [{Note8()}]{Note8}%
  \BibitemOpen
  \bibinfo {note} {The longitudinal and perpendicular parts of a vector field $\protect \mathbf {v}(\protect \mathbf {r})$ are most easily obtained in $\protect \mathbf {k}$-space \cite {cohen1997photons}, from its Fourier transform $\protect \mathbf {v}(\protect \mathbf {k})$: $\protect \mathbf {v}^\parallel (\protect \mathbf {k})=\protect \bm {\kappa }[\protect \bm {\kappa }\cdot \protect \mathbf {v}(\protect \mathbf {k})]$, and $\protect \mathbf {v}^\perp (\protect \mathbf {k})= \protect \mathbf {v}(\protect \mathbf {k})- \protect \mathbf {v}^\parallel (\protect \mathbf {k})$.}\BibitemShut {Stop}%
\bibitem [{Note9()}]{Note9}%
  \BibitemOpen
  \bibinfo {note} {For example, for a polarization density defined as in Eq.~\protect \eqref {eq:poldens}, $\protect \hat {U}_\protect \mathrm {PZW}$ is actually not a unitary transformation. Also, in this formulation, infinities arise in the polarization self-energy term; this is ultimately due to taking the square of a delta function, which is ill-defined. This goes to show in this case it is important to realize that delta `functions' are actually distributions, and should be treated as such.}\BibitemShut {Stop}%
\bibitem [{Note10()}]{Note10}%
  \BibitemOpen
  \bibinfo {note} {This is straightforward to show using the Parseval--Plancherel identity, and the aforementioned projection of the perpendicular component of a vector field in $\protect \mathbf {k}$-space \cite {cohen1997photons}.}\BibitemShut {Stop}%
\bibitem [{\citenamefont {Taylor}\ \emph {et~al.}(2020)\citenamefont {Taylor}, \citenamefont {Mandal}, \citenamefont {Zhou},\ and\ \citenamefont {Huo}}]{taylor2020resolution}%
  \BibitemOpen
  \bibfield  {author} {\bibinfo {author} {\bibfnamefont {M.~A.}\ \bibnamefont {Taylor}}, \bibinfo {author} {\bibfnamefont {A.}~\bibnamefont {Mandal}}, \bibinfo {author} {\bibfnamefont {W.}~\bibnamefont {Zhou}},\ and\ \bibinfo {author} {\bibfnamefont {P.}~\bibnamefont {Huo}},\ }\bibfield  {title} {\bibinfo {title} {Resolution of gauge ambiguities in molecular cavity quantum electrodynamics},\ }\href@noop {} {\bibfield  {journal} {\bibinfo  {journal} {Phys. Rev. Lett.}\ }\textbf {\bibinfo {volume} {125}},\ \bibinfo {pages} {123602} (\bibinfo {year} {2020})}\BibitemShut {NoStop}%
\bibitem [{\citenamefont {Stokes}\ and\ \citenamefont {Nazir}(2022)}]{stokes2022implications}%
  \BibitemOpen
  \bibfield  {author} {\bibinfo {author} {\bibfnamefont {A.}~\bibnamefont {Stokes}}\ and\ \bibinfo {author} {\bibfnamefont {A.}~\bibnamefont {Nazir}},\ }\bibfield  {title} {\bibinfo {title} {Implications of gauge freedom for nonrelativistic quantum electrodynamics},\ }\href@noop {} {\bibfield  {journal} {\bibinfo  {journal} {Rev. Mod. Phys.}\ }\textbf {\bibinfo {volume} {94}},\ \bibinfo {pages} {045003} (\bibinfo {year} {2022})}\BibitemShut {NoStop}%
\bibitem [{\citenamefont {Castagnola}\ \emph {et~al.}(2024{\natexlab{a}})\citenamefont {Castagnola}, \citenamefont {Riso}, \citenamefont {Barlini}, \citenamefont {Ronca},\ and\ \citenamefont {Koch}}]{castagnola2024polaritonic}%
  \BibitemOpen
  \bibfield  {author} {\bibinfo {author} {\bibfnamefont {M.}~\bibnamefont {Castagnola}}, \bibinfo {author} {\bibfnamefont {R.~R.}\ \bibnamefont {Riso}}, \bibinfo {author} {\bibfnamefont {A.}~\bibnamefont {Barlini}}, \bibinfo {author} {\bibfnamefont {E.}~\bibnamefont {Ronca}},\ and\ \bibinfo {author} {\bibfnamefont {H.}~\bibnamefont {Koch}},\ }\bibfield  {title} {\bibinfo {title} {Polaritonic response theory for exact and approximate wave functions},\ }\href@noop {} {\bibfield  {journal} {\bibinfo  {journal} {Wiley Interdiscip. Rev. Comput. Mol. Sci.}\ }\textbf {\bibinfo {volume} {14}},\ \bibinfo {pages} {e1684} (\bibinfo {year} {2024}{\natexlab{a}})}\BibitemShut {NoStop}%
\bibitem [{\citenamefont {Castagnola}\ \emph {et~al.}(2024{\natexlab{b}})\citenamefont {Castagnola}, \citenamefont {Haugland}, \citenamefont {Ronca}, \citenamefont {Koch},\ and\ \citenamefont {Schafer}}]{castagnola2024collective}%
  \BibitemOpen
  \bibfield  {author} {\bibinfo {author} {\bibfnamefont {M.}~\bibnamefont {Castagnola}}, \bibinfo {author} {\bibfnamefont {T.~S.}\ \bibnamefont {Haugland}}, \bibinfo {author} {\bibfnamefont {E.}~\bibnamefont {Ronca}}, \bibinfo {author} {\bibfnamefont {H.}~\bibnamefont {Koch}},\ and\ \bibinfo {author} {\bibfnamefont {C.}~\bibnamefont {Schafer}},\ }\bibfield  {title} {\bibinfo {title} {Collective strong coupling modifies aggregation and solvation},\ }\href@noop {} {\bibfield  {journal} {\bibinfo  {journal} {J. Phys. Chem. Lett.}\ }\textbf {\bibinfo {volume} {15}},\ \bibinfo {pages} {1428} (\bibinfo {year} {2024}{\natexlab{b}})}\BibitemShut {NoStop}%
\bibitem [{\citenamefont {Huang}\ and\ \citenamefont {Liang}(2025)}]{huang2025local}%
  \BibitemOpen
  \bibfield  {author} {\bibinfo {author} {\bibfnamefont {X.}~\bibnamefont {Huang}}\ and\ \bibinfo {author} {\bibfnamefont {W.}~\bibnamefont {Liang}},\ }\bibfield  {title} {\bibinfo {title} {Local molecular properties of non-interacting and hydrogen-bonded water systems in vibrational strong coupling regimes},\ }\href@noop {} {\bibfield  {journal} {\bibinfo  {journal} {ChemRxiv. 2025; doi:10.26434/chemrxiv-2025-kmz9j}\ } (\bibinfo {year} {2025})}\BibitemShut {NoStop}%
\bibitem [{\citenamefont {Milonni}(1994)}]{milonni2013quantum}%
  \BibitemOpen
  \bibfield  {author} {\bibinfo {author} {\bibfnamefont {P.~W.}\ \bibnamefont {Milonni}},\ }\href@noop {} {\emph {\bibinfo {title} {The quantum vacuum: {An} introduction to quantum electrodynamics}}}\ (\bibinfo  {publisher} {Academic press},\ \bibinfo {year} {1994})\BibitemShut {NoStop}%
\bibitem [{\citenamefont {Medina}\ \emph {et~al.}(2021)\citenamefont {Medina}, \citenamefont {Garc{\'\i}a-Vidal}, \citenamefont {Fern{\'a}ndez-Dom{\'\i}nguez},\ and\ \citenamefont {Feist}}]{medina2021few}%
  \BibitemOpen
  \bibfield  {author} {\bibinfo {author} {\bibfnamefont {I.}~\bibnamefont {Medina}}, \bibinfo {author} {\bibfnamefont {F.~J.}\ \bibnamefont {Garc{\'\i}a-Vidal}}, \bibinfo {author} {\bibfnamefont {A.~I.}\ \bibnamefont {Fern{\'a}ndez-Dom{\'\i}nguez}},\ and\ \bibinfo {author} {\bibfnamefont {J.}~\bibnamefont {Feist}},\ }\bibfield  {title} {\bibinfo {title} {Few-mode field quantization of arbitrary electromagnetic spectral densities},\ }\href@noop {} {\bibfield  {journal} {\bibinfo  {journal} {Phys. Rev. Lett.}\ }\textbf {\bibinfo {volume} {126}},\ \bibinfo {pages} {093601} (\bibinfo {year} {2021})}\BibitemShut {NoStop}%
\bibitem [{\citenamefont {S{\'a}nchez-Barquilla}\ \emph {et~al.}(2022)\citenamefont {S{\'a}nchez-Barquilla}, \citenamefont {Garc{\'\i}a-Vidal}, \citenamefont {Fern{\'a}ndez-Dom{\'\i}nguez},\ and\ \citenamefont {Feist}}]{sanchez2022few}%
  \BibitemOpen
  \bibfield  {author} {\bibinfo {author} {\bibfnamefont {M.}~\bibnamefont {S{\'a}nchez-Barquilla}}, \bibinfo {author} {\bibfnamefont {F.~J.}\ \bibnamefont {Garc{\'\i}a-Vidal}}, \bibinfo {author} {\bibfnamefont {A.~I.}\ \bibnamefont {Fern{\'a}ndez-Dom{\'\i}nguez}},\ and\ \bibinfo {author} {\bibfnamefont {J.}~\bibnamefont {Feist}},\ }\bibfield  {title} {\bibinfo {title} {Few-mode field quantization for multiple emitters},\ }\href@noop {} {\bibfield  {journal} {\bibinfo  {journal} {Nanophotonics}\ }\textbf {\bibinfo {volume} {11}},\ \bibinfo {pages} {4363} (\bibinfo {year} {2022})}\BibitemShut {NoStop}%
\bibitem [{\citenamefont {Chuang}\ and\ \citenamefont {Hsu}(2024)}]{chuang2024microscopic}%
  \BibitemOpen
  \bibfield  {author} {\bibinfo {author} {\bibfnamefont {Y.-T.}\ \bibnamefont {Chuang}}\ and\ \bibinfo {author} {\bibfnamefont {L.-Y.}\ \bibnamefont {Hsu}},\ }\bibfield  {title} {\bibinfo {title} {Microscopic theory of exciton--polariton model involving multiple molecules: Macroscopic quantum electrodynamics formulation and essence of direct intermolecular interactions},\ }\href@noop {} {\bibfield  {journal} {\bibinfo  {journal} {J. Chem. Phys.}\ }\textbf {\bibinfo {volume} {160}} (\bibinfo {year} {2024})}\BibitemShut {NoStop}%
\bibitem [{\citenamefont {Stone}(2013)}]{stone2013theory}%
  \BibitemOpen
  \bibfield  {author} {\bibinfo {author} {\bibfnamefont {A.}~\bibnamefont {Stone}},\ }\href@noop {} {\emph {\bibinfo {title} {The theory of intermolecular forces}}},\ \bibinfo {edition} {2nd}\ ed.\ (\bibinfo  {publisher} {Oxford University Press},\ \bibinfo {year} {2013})\BibitemShut {NoStop}%
\end{thebibliography}%
\end{document}